\documentclass[12pt,draftclsnofoot,onecolumn]{IEEEtran}
\usepackage{amsmath,amsfonts}
\usepackage[ruled]{algorithm2e}
\usepackage{array}
\usepackage[caption=false,font=normalsize,labelfont=sf,textfont=sf]{subfig}
\usepackage{textcomp}
\usepackage{stfloats}
\usepackage{url}
\usepackage{amssymb}
\usepackage{verbatim}
\usepackage{graphicx}
\usepackage{cuted}
\usepackage{bm} 
\usepackage{multirow}
\newcommand{\tabincell}[2]{\begin{tabular}{@{}#1@{}}#2\end{tabular}} 

\usepackage{cite}
\begin{document}

\title{\LARGE{\textbf{Diagonally Reconstructed Channel Estimation for MIMO-AFDM with Inter-Doppler Interference in Doubly Selective Channels}}}

\author{Haoran Yin, Xizhang Wei, Yanqun Tang, and Kai Yang, \textit{Member, IEEE}
	\vspace{-1.5em}
	\thanks{
		This work was supported by Guangdong Natural Science Foundation
		under Grant 2019A1515011622. (\textit{Corresponding author: Yanqun Tang.})
		
		 Haoran Yin, Xizhang Wei, and Yanqun Tang are with the School of Electronics and Communication Engineering, Sun Yat-sen University, China (e-mail: yinhr6@mail2.sysu.edu.cn, weixzh7@mail.sysu.edu.cn, tangyq8@mail.sysu.edu.cn).
		
		Kai Yang is with the School of Information and Electronics, Beijing Institute of Technology, Beijing, China (email: yangkai@ieee.org).
		
		}  
}
\markboth{}%
{Shell \MakeLowercase{\textit{et al.}}: A Sample Article Using IEEEtran.cls for IEEE Journals}
\maketitle
\vspace{-2.5em}

\begin{abstract}
	\vspace{-0.5em}
\textbf{On the heels of orthogonal time frequency space (OTFS) modulation, the recently discovered affine frequency division multiplexing (AFDM) is a promising waveform for the sixth-generation wireless network. In this paper, we study the  widely-used embedded pilot-aided (EPA) channel estimation in multiple-input multiple-output AFDM (MIMO-AFDM) system with fractional Doppler shifts. 
We first formulate the vectorized input-output relationship of MIMO-AFDM, and theoretically prove that MIMO-AFDM can achieve full diversity in doubly selective channels. Then we illustrate the implementation of EPA channel estimation in MIMO-AFDM and unveil that serious inter-Doppler interference (IDoI) occurs if we try to estimate the 
channel gain, delay shift, and Doppler shift of each propagation path. To address this issue, the diagonal reconstructability of AFDM subchannel matrix is studied and a low-complexity embedded pilot-aided diagonal reconstruction (EPA-DR) channel estimation scheme is proposed. The EPA-DR scheme calculates the AFDM effective channel matrix directly without estimating the three channel parameters, eliminating the severe IDoI inherently. Since the effective channel matrix is necessary for MIMO-AFDM receive processing, we believe this is an important step to bring
AFDM towards practical communication systems. Finally, we investigate the orthogonal resource allocation of affine frequency division multiple access (AFDMA) system. Simulation results validate the effectiveness of the proposed EPA-DR scheme.}
\end{abstract}
\vspace{-0.5em}

\section{Introduction}
The sixth-generation wireless network is envisioned to provide ultra-reliable, high data rate, and low-latency communications in high-mobility scenarios, including  vehicle-to-vehicle (V2V), unmanned aerial vehicles (UAV), high-speed trains, and low-earth-orbit satellite (LEOS), etc. The dynamic channels therein
are characterized by heavy delay-Doppler spreads, which cast a huge challenge to the current widely adopted waveforms, such as orthogonal frequency division multiplexing (OFDM). The non-negligible Doppler shifts greatly devastate  the orthogonality between the subcarriers in OFDM, which is a serious problem especially in the case of higher frequency bands used in the future communication systems \cite{bb22.10.24.2, bb22.10.29.5}. Therefore, many efforts have been made to design a new modulation waveform to accommodate time- and frequency-selective channels. In particular, a two-dimensional (2D) modulation waveform named orthogonal time frequency space (OTFS) has attracted substantial attention \cite{bb2, bb22.10.29.3}. The main idea of OTFS is multiplexing the information symbols in the delay-Doppler (DD) domain, where a quasi-static channel representation can be obtained. The recent works on OTFS have shown the performance superiority of OTFS over OFDM in both single-antenna and multiple-antenna systems \cite{bb12, bb4, b4, bb15, bb22.10.29.2,bb22.11.17.1,bb105,bb106,bb22.10.29.1, bb23.4.8.1}. However, a disadvantage of OTFS is the heavy guard symbol overhead when conducting the widely used embedded pilot-aided (EPA) channel estimation due to its 2D structure \cite{bb6}.

Another promising candidate is affine frequency division multiplexing (AFDM), which is a newly discovered waveform and always attains full diversity in doubly selective channels \cite{bb5}. Information symbols in AFDM are multiplexed on a set of orthogonal chirps via inverse \emph{discrete affine Fourier transform} (DAFT), which is a generalization of the widely adopted inverse \emph{discrete Fourier transform} (DFT) \cite{bb7,bb8}. It is shown in \cite{bb5} that the AFDM modulation/demodulation can be  implemented efficiently using the OFDM modulator/demodulator as an inner core. Moreover, by tuning the chirp slope according to the Doppler profile of the channel appropriately, AFDM manages to separate the propagation paths with distinct delay or Doppler shifts in the one-dimensional DAFT domain. Hence, similar to OTFS, AFDM can transform a time-variant channel in
the time-frequency domain into a  quasi-time-invariant channel. Some recent works are conducted to unleash the potential of AFDM. For example, by exploring the diagonal property of the AFDM effective channel matrix, a weighted maximal-ratio combining based low-complexity iterative decision feedback equalizer and a low-complexity MMSE detector were proposed for AFDM signal detection in \cite{bb102}. The potential of AFDM for communication at high-frequency bands was presented in \cite{bb101},  showing that AFDM is robust against radio frequency impairments, such as carrier frequency offset and phase noise. The suitability of AFDM for integrated sensing and communications was investigated in \cite{bb2022.11.11.1} and \cite{bb2022.11.11.2}.
Simulation results in \cite{bb5,bb101, bb102, bb6} show that AFDM has a bit error ratio (BER) performance similar to OTFS and outperforms OFDM greatly in doubly selective channels. 

While the aforementioned excellent works have revealed the superiority of AFDM, they all assume perfect channel state information (CSI), which has to be estimated in practice. A pilot-aided path detection based channel estimation scheme was presented in \cite{bb22.11.13.1}, where the delay and Doppler shift of each propagation path was assumed to be the integral multiple of the sample interval and subcarrier spacing, respectively. These assumptions are partly rational as generally a wideband system can provide sufficient delay resolution so that fractional delay shifts do not need to be considered \cite{b4}. However, the Doppler resolution depends on the time duration of AFDM frame, which is typically small due to the low-latency demand and channel aging problem in high-mobility scenarios. Assuming integer Doppler shifts in the condition of insufficient Doppler resolution will induce significant modelling errors \cite{bb13}. To tackle this problem, an embedded pilot-aided approximated maximum likelihood (EPA-AML) channel estimation scheme was proposed for single-input single-output AFDM (SISO-AFDM) in \cite{bb6}, where the fractional Doppler was taken into account preliminarily. However, the authors in \cite{bb6} hypothesized that the delay shift of each propagation path is different from each other and the number of propagation paths is known in advance. These hypotheses undermine its feasibility greatly given that the delay-Doppler profile of any general doubly selective channel should cover the three cases listed in Table \ref{tabel7-31-1}. Moreover, high computation complexity is required in EPA-AML, making it not suitable for multiple-input multiple-output AFDM (MIMO-AFDM) system.
\begin{table}[ht]
	\vspace{-2em}
	\renewcommand\arraystretch{1}
	\centering
	\caption{Three Cases of Delay-Doppler Profile in Doubly Selective Channel}
	\label{tabel7-31-1}
	\vspace{-1.5em}
	\begin{tabular}{|c|c|}
		\hline
		Case 1  & Paths with different delay shifts and different Doppler shifts\\ 
		\hline
		Case 2 & Paths with different delay shifts and same Doppler shift \\
		\hline
		\textbf{Case 3} & Paths with same delay shift and different Doppler shifts \\			
		\hline
	\end{tabular}
\vspace{-1.5em}
\end{table}

In this paper, we aim to address the issue of MIMO-AFDM channel estimation with specific attention to fractional Doppler shifts. To this end, we investigate the influence of fractional Doppler shifts on the EPA channel estimation in MIMO-AFDM. In specific, we reveal that there exists serious inter Doppler interference (IDoI) among the received pilot symbols when two propagation paths have the same delay shift and different Doppler shifts. This casts a huge challenge to the estimation of the three channel parameters of each propagation path, i.e., the channel gain, the delay shift, and the Doppler shift. In order to avoid the IDoI, we prove theoretically that the AFDM subchannel matrix is diagonally reconstructable, making it possible to estimate the AFDM effective channel matrix directly rather than first estimate the three channel parameters and then calculate it. Furthermore, we unveil that exploring the diagonal reconstructability to perform channel estimation can inherently avoid the severe IDoI. In contrast to the state-of-the-art method, our proposed channel estimation scheme is more robust to the fractional Doppler and enjoys much lower computation complexity. The BER of the MIMO-AFDM with perfect CSI and estimated CSI are compared to demonstrate the
effectiveness of the proposed algorithm.
Our contributions can be summarized as follows.
\begin{itemize} 
	\item[$\bullet$]
	We develop the settings of MIMO-AFDM system and derive the associated input-output relationship in the DAFT domain. Based on that, we extend the diversity analysis of SISO-AFDM in \cite{bb6} to MIMO-AFDM with perfect CSI. It is shown that MIMO-AFDM can achieve full diversity in doubly selective channels, where the full diversity order refers to the number of receive antennas multiplied by the number of paths that are separable in either delay or Doppler domain. 
\end{itemize}
\begin{itemize} 
	\item[$\bullet$]
	After demonstrating the implementation of EPA channel estimation in MIMO-AFDM, we study the interference among the received pilot symbols caused by fractional Doppler. We reveal that there are four kinds of interference, namely inter-Doppler interference (IDoI), inter-delay interference (IDI), inter-pilot interference (IPI), and inter-pilot-data interference (IPDI). Among them, the IDoI is the most serious and impossible to be eliminated if we try to estimate the channel gain, delay shift, and Doppler shift of each propagation path.
\end{itemize}
\begin{itemize} 
	\item[$\bullet$]
	Next, we prove that the AFDM subchannel matrix can be diagonally reconstructed by utilizing some precalculable transform factors iteratively.  Based on this unique characteristic, we propose a low-complexity channel estimation method named embedded pilot-aided diagonal reconstruction (EPA-DR), which can avoid IDoI perfectly.
\end{itemize}
\begin{itemize} 
	\item[$\bullet$]
	Finally, we propose an orthogonal resource allocation scheme for affine frequency division multiple access (AFDMA) system  based on the EPA-DR channel estimation method. By exploring the differences in the delay-Doppler profiles between the base station and all the users, we manage to reduce the guard symbol overhead to the greatest extent.
\end{itemize}

The rest of this paper is organized as follows. Section \ref{sec2} reviews the basic concepts of AFDM and introduces MIMO-AFDM system, which lays the foundations for the interference analysis of the EPA channel estimation in Section \ref{sec3}. Section \ref{sec4} presents the EPA-DR channel estimation method for MIMO-AFDM, followed by the resource allocation for AFDMA in Section \ref{sec5}. Simulation results are presented in Section \ref{secResults}, while Section \ref{sec6} concludes this paper.

\textit{Notations:} Symbols $\mathbf{a}[i]$ and $\mathbf{A}[i,j]$ denote the $i$-th element of $\mathbf{a}$ and the ($i, j$)-th element of $\mathbf{A}$ respectively; $\mathbb{C}$ denotes the set of complex numbers and $\mathbb{C}^{M \times N}$ denotes the set of all $M \times N$ matrices with complex entries;  $\mathbf{I}_{N}$ denotes the identity matrix
of size $N \times N$;
$\mathbf{a} \sim \mathcal{C} \mathcal{N}\left(\mathbf{0}, N_{0} \mathbf{I}_{N}\right)$ means that $\mathbf{a}$ follows the complex Gaussian distribution with
zero mean and covariance $N_{0} \mathbf{I}_{N}$; 
$\delta(\cdot)$ denotes the Dirac delta function; $\operatorname{diag}(\cdot)$ denotes a square diagonal matrix with the elements of input vector on the main diagonal; $(\cdot)^{*}$, $(\cdot)^{\mathrm{H}}$, $(\cdot)^{T}$, and $\left\| \cdot \right\|$ denote the conjugate,
Hermitian, transpose, and Euclidean norm operations; $\lvert \cdot \rvert$ denotes the absolute value of a
complex scalar; $(\cdot)_{N}$ denotes the modulus operation with respect to $N$;
$\mathbb{E}[\cdot]$ denotes  the expectation; $Q(\cdot)$ denotes the tail distribution
function of the standard normal distribution.

\vspace{-0.5em}
\section{AFDM System Model}
\label{sec2}
In this section, the basic concepts of AFDM from \cite{bb6}  and  \cite{bb5} are reviewed. Based on that, we establish the settings of MIMO-AFDM system.

\begin{figure}[htbp]
	\centering
	\vspace{-1em}
	\includegraphics[width=0.79\textwidth,height=0.11\textwidth]{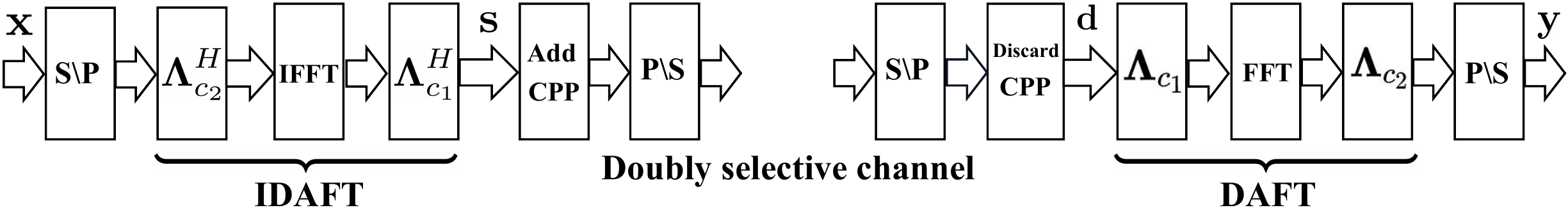}
	\vspace{-1.8em}
	\caption{SISO-AFDM modulation/demodulation block diagrams \cite{bb6}.}
	\label{2-1}
	\vspace{-0.9em}
\end{figure}

\vspace{-1.5em}
\subsection{AFDM modulation}
\emph{\textbf{1) AFDM modulation}}: Fig. \ref{2-1} shows the modulation/demodulation block diagrams of SISO-AFDM system. Let $T_{s}$ denote sample period (delay resolution), $N$ denote the number of subcarriers (chirps), then AFDM signal has a bandwidth $B=\frac{1}{T_{s}}$, subcarrier spacing (Doppler resolution) $\Delta f=\frac{B}{N}=\frac{1}{NT_{s}}$.
Let $\mathbf{x} \in \mathbb{A}^{N \times 1}$ denote a vector of $N$ quadrature amplitude modulation (QAM) symbols that reside on the DAFT domain, $\mathbb{A}$ represents the modulation alphabet. After the serial to parallel operation, $N$-point inverse DAFT (IDAFT) is performed to map $\mathbf{x}$ to the time domain as \cite{bb6}
\begin{equation}
	\vspace{-0.2em}
	s[n]= \sum_{m=0}^{N-1} x[m] \phi_{m}[n],  \ n=0, \cdots, N-1
	\label{eq2-1}
\end{equation}
where $n$ and $m$ denote the time and DAFT domains indices, respectively, and subcarrier $\phi_{m}[n]$ is given by
\begin{equation} \phi_{m}[n]=\frac{1}{\sqrt{N}} e^{j 2 \pi\left(c_{1} n^{2}+c_{2} m^{2}+  n m / N\right)}, \ m=0, \cdots, N-1
\end{equation}
$c_{1}$ and $c_{2}$ are two AFDM parameters, where $c_{1}$ determines the chirps' slope. Equation (\ref{eq2-1}) can be written in matrix form as
\begin{equation}
\mathbf{s} =  \boldsymbol{\Lambda}_{c_{1}}^{H} \mathbf{F}^{H} \boldsymbol{\Lambda}_{c_{2}}^{H} \mathbf{x} = \mathbf{A}^{H} \mathbf{x}
\label{eq2-2}
\end{equation}
where $\mathbf{A} = \boldsymbol{\Lambda}_{c_{2}} \mathbf{F} \boldsymbol{\Lambda}_{c_{1}}\in\mathbb{C}^{N\times N}$ represents the DAFT matrix,  $\mathbf{F}$ is the DFT matrix with entries $e^{-j 2 \pi m n / N} / \sqrt{N}$, $\boldsymbol{\Lambda}_{c}\triangleq\operatorname{diag}\left(e^{-j 2 \pi c n^{2}}, n=0,1, \ldots,   N-1\right)$. Before transmitting $\mathbf{s}$, an \emph{chirp-periodic} prefix (CPP) should be added, which plays the same role as the cyclic prefix (CP) in OFDM to cope with the multipath propagation and makes the channel lie in a periodic domain equivalently.

\subsection{Channel model}
Assume the doubly selective channel has the following impulse response  with delay $\tau$ and Doppler $\kappa$ as
\begin{equation}
	h(\tau, \kappa)=\sum_{i=1}^{P} h_{i} \delta\left(\tau-l_{i}T_{s}\right) \delta\left(\kappa-\nu_{i}\Delta f\right)
	\label{eq2-4}
\end{equation}
where $P$ is the number of paths, $h_{i}$ denotes the channel gain of the $i$-th path, non-negative integer
$l_{i} \in [0, l_{\max}]$ is the associated delay
normalized with $T_{s}$,  $\nu_{i}=\alpha_{i}+\beta_{i}$ represents the associated Doppler shift normalized with subcarrier spacing and has a finite support bounded by $[-\nu_{\max}, \nu_{\max}]$, $\alpha_{i} \in [-\alpha_{\max}, \alpha_{\max}]$ and $\beta_{i} \in (-\frac{1}{2}, \frac{1}{2}]$ are the integer and fractional parts of $\nu_{i}$ respectively, $\nu_{\max}$ denotes the maximum Doppler and $\alpha_{\max}$ denotes its integer component. In this paper, we assume that 
$l_{\max}$ and $\nu_{\max}$ are known in advance.

\subsection{AFDM demodulation}
At the receiver,  the relationship between the received time domain symbols $\mathbf{d}$ and $\mathbf{s}$ can be expressed as 
\begin{equation}
	d[n]=\sum_{i=1}^{P} h_{i} e^{-j  \frac{2 \pi}{N}\nu_{i}n} s[(n-l_{i})_{ N}]+v[n]
	\label{eq2-5}
\end{equation}
where  $v \sim \mathcal{C} \mathcal{N}\left(0, N_{0}\right)$ represents the additive white gaussian noise (AWGN) component. Equation (\ref{eq2-5}) can be vectorized as
\begin{equation}
 	\mathbf{d} = \sum_{i=1}^{P} h_{i} \mathbf{\tilde{H}}_{i}  \mathbf{s}  + \mathbf{v}= \mathbf{\tilde{H}} \mathbf{s} + \mathbf{v}  
 \label{eq2-6}
\end{equation}
where $\mathbf{v} \sim \mathcal{C} \mathcal{N}\left(\mathbf{0}, N_{0} \mathbf{I}_{N}\right)$ is the time domain noise vector, $\mathbf{\tilde{H}}\in\mathbb{C}^{N\times N}$ denotes the effective time domain channel matrix, $\mathbf{\tilde{H}}_{i}= \boldsymbol{\Delta}_{\nu_{i}} \boldsymbol{\Pi}^{l_{i}}$ represents the time domain subchannel matrix of the $i$-th path (each path can be viewed as one subchannel), $\boldsymbol{\Pi}$ denotes the forward cyclic-shift matrix which models the delay, while the digital frequency shift matrix $\boldsymbol{\Delta}_{\nu_{i}} \triangleq \operatorname{diag}\left(e^{-j \frac{2 \pi}{N} \nu_{i} n}, n=0,1, \cdots, N-1\right)$ models the Doppler. 

Finally, $N$-point DAFT is implemented and  $\mathbf{d}$ are transformed to the DAFT domain symbols $\mathbf{y}$ with
\begin{equation}
	y[m]=\sum_{m=0}^{N-1} d[n] \phi_{m}^{*}[n]   + w[m] 
	\label{eq2-7}
\end{equation}
where $w$ represents the noise in the DAFT domain. The matrix representation of (\ref{eq2-7}) is 
\begin{equation}
\mathbf{y} = \boldsymbol{\Lambda}_{c_{2}} \mathbf{F} \boldsymbol{\Lambda}_{c_{1}} \mathbf{d} = \mathbf{A} \mathbf{d}.
\label{eq2-8}
\end{equation}
Since $\mathbf{A}$ is a unitary matrix, $w$ has the same statistical properties as $v$.

\subsection{Input-output relationship}
\label{sec2-4}
Substituting  (\ref{eq2-1}) and (\ref{eq2-5}) into (\ref{eq2-7}), we have the input-output relationship of SISO-AFDM system  in the DAFT domain as \cite{bb6}
\begin{equation}
	\label{eq2-9}
	y[m]  = \frac{1}{N}\sum_{i=1}^{P}\sum_{m'=0}^{N-1}   
	h_{i} 	\underbrace{e^{j \frac{2 \pi}{N}\left(Nc_{1} l_{i}^{2}-m' l_{i}+N c_{2}\left(m'^{2}-m^{2}\right)\right)}}_{ \text{$\mathcal{C}(l_{i}, m, m') $}} 
	\underbrace{\frac{e^{j 2 \pi\left(m+\operatorname{ind}_{i}-m'+\beta_{i}\right)}-1}{e^{j \frac{2 \pi}{N}\left(m+\operatorname{ind}_{i}-m'+\beta_{i}\right)}-1}}_{\text{Spreading factor $	\mathcal{F}(l_{i},\nu_{i}, m, m')$}}
	x[m']+w[m] 			
\end{equation}
where the index indicator of the $i$-th path is defined as
\begin{equation} 
	\operatorname{ind}_{i} \triangleq\left(\alpha_{i}+2 N c_{1} l_{i}\right)_{N}. 
	\label{eq7-3-3}
\end{equation}
For the convenience of illustration, we define
\begin{equation}  
\mathcal{C}(l_{i}, m, m') = e^{j \frac{2 \pi}{N}\left(Nc_{1} l_{i}^{2}-m' l_{i}+N c_{2}\left(m'^{2}-m^{2}\right)\right)}
\label{eq2022.11.09} 
\end{equation}
and the spreading factor caused by fractional Doppler as
\begin{equation}
	\mathcal{F}(l_{i},\nu_{i}, m, m')= \frac{e^{j 2 \pi\left(m+\operatorname{ind}_{i}-m'+\beta_{i}\right)}-1}{e^{j \frac{2 \pi}{N}\left(m+\operatorname{ind}_{i}-m'+\beta_{i}\right)}-1}. 
	\label{eq2-11-3}
\end{equation}

The matrix form of (\ref{eq2-9}) can be obtained by substituting (\ref{eq2-2}) and  (\ref{eq2-6}) into (\ref{eq2-8}) as
\begin{equation}
	\label{eq2-12-2}
	\mathbf{y} = \sum_{i=1}^{P} h_{i} \mathbf{H}_{i} \mathbf{x} + \mathbf{w} = \mathbf{H}_{\text{eff}}\mathbf{x} + \mathbf{w}
\end{equation}
where $\mathbf{H}_{i} = \mathbf{A} \mathbf{\tilde{H}}_{i} \mathbf{A}^{H}$ denotes the DAFT domain subchannel matrix of the $i$-th path, $\mathbf{H}_{\text{eff}} = \sum_{i=1}^{P} h_{i} \mathbf{H}_{i}$ is the effective channel matrix, $\mathbf{w}\sim \mathcal{C} \mathcal{N}\left(\mathbf{0}, N_{0} \mathbf{I}_{N}\right)$ is the DAFT domain noise vector. Note that
\begin{equation}
	\label{eq2-13-2}
	\mathbf{H}_{i}[m,m'] = \frac{1}{N}\mathcal{C}(l_{i}, m, m')\mathcal{F}(l_{i},\nu_{i}, m, m').
\end{equation}

\emph{\textbf{Remark 1:}} It has been proven in \cite{bb6} (Theorem 1) that AFDM can achieve full diversity in doubly selective channels as long as
\begin{equation} 
c_{1} = \frac{2 (\alpha_{\max }+k_{\nu})+1}{2N}
\label{eq7-03-1}
\end{equation}
and $c_{2}$ is set as either an arbitrary irrational number or a rational number sufficiently smaller than $\frac{1}{2N}$ (spacing factor $k_{\nu}$ is a non-negative integer used to combat the fractional Doppler and will be illustrated in Section \ref{sec3}).

\vspace{-0.5em}
\subsection{MIMO-AFDM System}
\begin{figure}[htbp]
	\centering
	\vspace{-1.3em}
	\includegraphics[width=0.780\textwidth,height=0.20\textwidth]{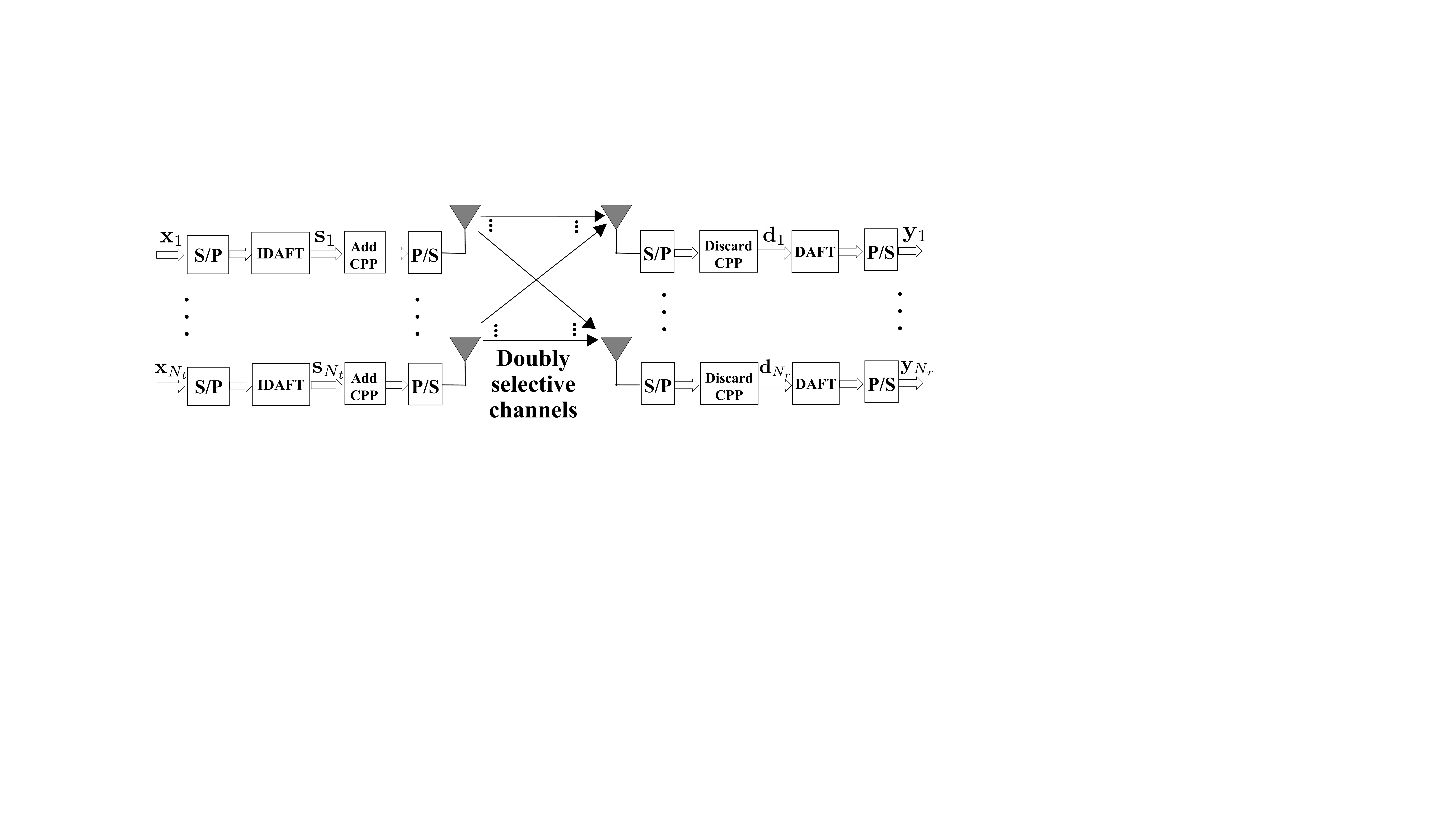}
	\vspace{-1.5em}
	\caption{$N_{t} \times N_{r}$ MIMO-AFDM modulation/demodulation block diagrams.}
	\label{2-2}
	\vspace{-0.9em}
\end{figure}
We next introduce MIMO-AFDM system with its modulation/demodulation block diagrams provided in Fig. \ref{2-2}. Let $N_{t}$ and $N_{r}$ denote the number of transmit antennas (TA) and receive antennas (RA) respectively. Then the linear system model based input-output relationship between the $r$-th RA and $N_{t}$ TAs from an $N_{t} \times N_{r}$ MIMO-AFDM system can be derived from (\ref{eq2-9}) as 
\begin{equation}
	\label{eq2-111}
	y_{r}[m]=
	\sum_{t=1}^{N_{t}}
	\sum_{i=1}^{P}
	\sum_{m'=0}^{N-1} 	 
	\frac{1}{N}h_{i}^{[r,t]} 
\mathcal{C}(l_{i}, m, m') \mathcal{F}(l_{i},\nu_{i}, m, m')
	x_{t}[m']+w_{r}[m] 
\end{equation}
where $ m \in [0,N-1]$, integer $r \in \left[1, N_{r}\right]$ and $t \in \left[1, N_{t}\right]$ denote the index of the RA and TA respectively, $h^{[r,t]}_{i}$ is the channel gain of the $i$-th
 path between the $r$-th RA and the $t$-th TA, $w_{r} \sim \mathcal{C} \mathcal{N}\left(0, N_{0}\right)$ represents the noise in DAFT domain at the $r$-th RA. Then the matrix form of the input-output relationships between all pairs of RAs and TAs can be denoted as
\begin{equation}
	\begin{aligned}
		\mathbf{y}_{1} &=\mathbf{H}_{1,1} \mathbf{x}_{1}+\mathbf{H}_{1,2} \mathbf{x}_{2}+\cdots+\mathbf{H}_{1,N_{t}} \mathbf{x}_{N_{t}}+\mathbf{w}_{1}  \\
		\vdots & \\
		\mathbf{y}_{N_{r}} &=\mathbf{H}_{N_{r}, 1} \mathbf{x}_{1}+\mathbf{H}_{N_{r}, 2} \mathbf{x}_{2}+\cdots+\mathbf{H}_{N_{r},N_{t}} \mathbf{x}_{N_{t}}+\mathbf{w}_{N_{r}}
	\end{aligned}
	\label{eq100}
\end{equation}
with 
\begin{equation}
 \mathbf{H}_{r,t}=\sum_{i=1}^{P} h_{i}^{[r,t]} \mathbf{H}_{i} 
 \label{eq7-15-1}
\end{equation}
representing the effective channel matrix between the $r$-th RA and the $t$-th TA, noise vector $\mathbf{w} \sim \mathcal{C} \mathcal{N}\left(\mathbf{0}, N_{0} \mathbf{I}_{N}\right)$. For the sake of compactedness, we define the effective MIMO channel matrix for the above MIMO-AFDM system as 
\vspace{-0.1em}
\begin{equation}
	\vspace{-0.1em}
	\mathbf{H}_{\text {MIMO}}=\left[\begin{array}{ccc}
		\mathbf{H}_{1,1} &  \ldots & \mathbf{H}_{1 ,N_{t}} \\
	
		\vdots &  \ddots & \vdots \\
		\mathbf{H}_{N_{r}, 1} &  \ldots & \mathbf{H}_{N_{r}, N_{t}}
	\end{array}\right]
	\label{eq.77}
\end{equation}
where $\mathbf{H}_{\text {MIMO}}\in\mathbb{C}^{NN_{r}\times NN_{t}}$,  transmitted vector $\mathbf{x}_{\text {MIMO}}=\left[\mathbf{x}_{1}^{T}, \mathbf{x}_{2}^{T},\cdots, \mathbf{x}_{N_{t}}^{T}\right]^{T}\in\mathbb{C}^{NN_{t}\times 1}$,
received vector $\mathbf{y}_{\text {MIMO}}=\left[\mathbf{y}_{1}^{T}, \mathbf{y}_{2}^{T}, \cdots, \mathbf{y}_{N_{r}}^{T}\right]^{T}\in\mathbb{C}^{NN_{r}\times 1}$, and noise vector
$\mathbf{w}_{\text {MIMO }}=\left[\mathbf{w}_{1}^{T}, \mathbf{w}_{2}^{T}, \cdots, \mathbf{w}_{N_{r}}^{T}\right]^{T}\in\mathbb{C}^{NN_{r}\times 1}$. Then (\ref{eq100}) can be rewritten as 
\vspace{-0.3em}
\begin{equation}
	\vspace{-0.3em}
	\mathbf{y}_{\text {MIMO}}=\mathbf{H}_{\text {MIMO}} \mathbf{x}_{\mathrm{MIMO}}+\mathbf{w}_{\text {MIMO}}.
	\label{eq.8}
\end{equation}

\emph{\textbf{Theorem 1:} For a linear time-varying channel with a maximum normalized delay $l_{\max}$ and maximum normalized Doppler $\alpha_{\max}$, MIMO-AFDM with parameter $c_{1}$ satisfying (\ref{eq7-03-1}) achieves full diversity order of $PN_{r}$.} 

\emph{Proof}: The proof is given in Appendix \ref{APP1}.

In the sequel, we study the EPA channel estimation of MIMO-AFDM with fractional Doppler shifts.

\vspace{-0.1em}
\section{Interference Analysis of EPA Channel Estimation in MIMO-AFDM}
\label{sec3}

In practice, $\mathbf{H}_{\text {MIMO}}$ must be estimated at the receiver to perform signal detection with (\ref{eq.8}). To meet this requirement, we analyze the biggest challenge of the widely-used EPA channel estimation in MIMO-AFDM system with fractional Doppler.

\begin{figure}[htbp]
	\vspace{-0.8em}
	\centering
	\includegraphics[width=0.4\textwidth,height=0.4\textwidth]{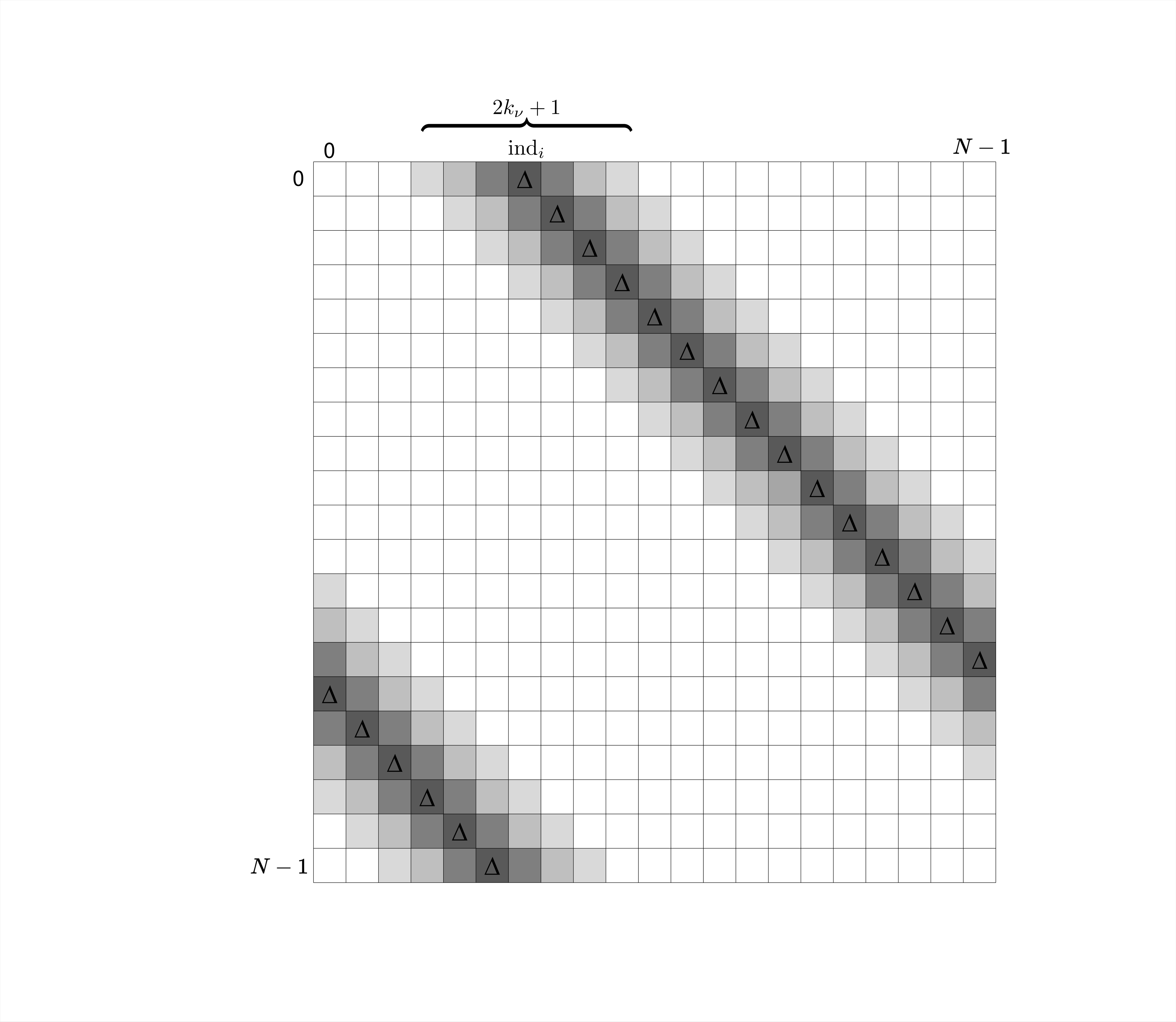}
	\vspace{-1.3em}
	\caption{Structure of $\mathbf{H}_{i}$ with fractional Doppler shift \cite{bb6} (the darker the color, the larger the magnitude; ‘$\Delta$’: the central point in each row or column).}
	\label{fig7-3-1}
	\vspace{-1.8em}
\end{figure}

\vspace{-0.8em}
\subsection{Effective channnel matrix analysis}
We first illustrate the influence of fractional Doppler on the AFDM subchannel matrix. From (\ref{eq2-13-2}), we have
\begin{equation}
	|\mathbf{H}_{i}[m,m']|=|\frac{1}{N}\mathcal{F}(l_{i},\nu_{i}, m, m')|.
	\label{eq.43-9}
\end{equation}
The magnitude $ | \mathcal{F}(l_{i},\nu_{i}, m, m') |$ reaches the peak at $m'=\left(m+\operatorname{ind}_{i}\right)_{N}$ and decreases as $m'$ moves away from $(m + \operatorname{ind}_{i})_{N}$. 
Therefore, in the $m$-th row of $\mathbf{H}_{i}$, $ m \in [0,N-1]$, the index of entry that has the greatest magnitude, i.e., \textbf{central point}, is 
\begin{equation}
 m'=(m+\operatorname{ind}_{i})_{N}. 
\label{eq7-5-1}
\end{equation}
While the energy of the adjacent entries decays as their indices move away from $(m+\operatorname{ind}_{i})_{N}$, i.e., the larger the value of $|m+ \operatorname{ind}_{i}-m'|$, the smaller magnitude of $\mathbf{H}_{i}[m,m']$. This also implies that the non-zero entries are “diagonally” distributed in $\mathbf{H}_{i}$, as shown in Fig. \ref{fig7-3-1}. Note that in the case of integer Doppler, there is only one non-zero entry (i.e., the central point) in each row of $\mathbf{H}_{i}$.  The “spreading" phenomenon in $\mathbf{H}_{i}$ shown in Fig. \ref{fig7-3-1} is caused by fractional Doppler shift \cite{bb6}, which casts a huge challenge to the CSI estimation and will be illustrated later in Section \ref{sec3-3}.

\begin{figure*}[htbp]
	\centering
	\vspace{-0.9em}
	\includegraphics[width=0.87\textwidth,height=0.37\textwidth]{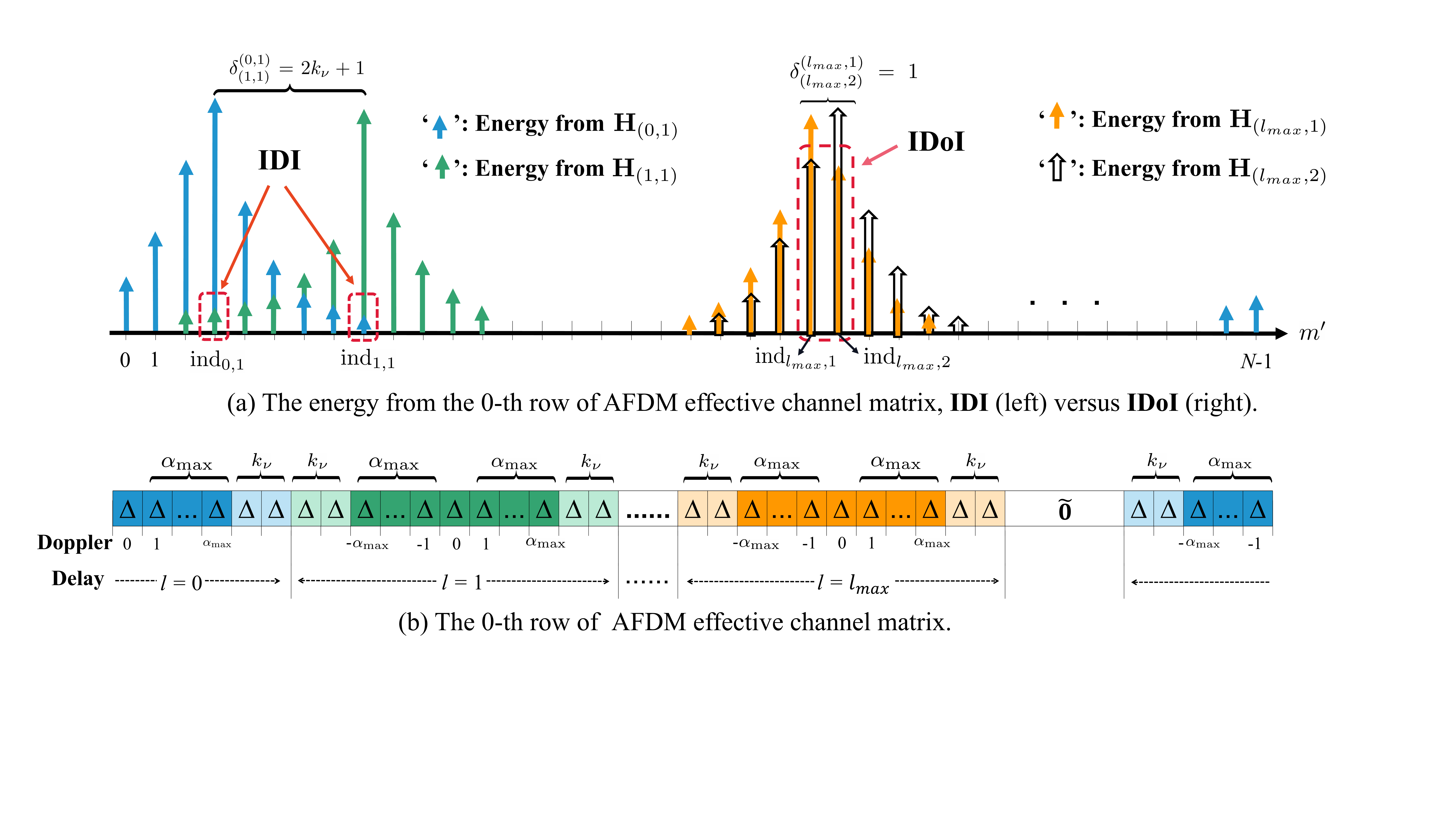}
	\vspace{-0.9em}
	\caption{Characteristics illustration of AFDM effective channel matrix (‘$\Delta$’: large non-zero value, $\tilde{\mathbf{0}}$: small non-zero value).}
	\label{sci4-3}
	\vspace{-1.8em}
\end{figure*}

Moreover, submitting $c_{1}$ in (\ref{eq7-03-1}) into the index indicator defined in (\ref{eq7-3-3}), we have 
\vspace{-0.2em}
\begin{equation}
	\vspace{-0.2em}
 \operatorname{ind}_{i}=\big(\alpha_{i}+(2(\alpha_{\max }+ k_{\nu})+1)l_{i}\big)_{N}. 
 	\label{eq7-3-2}
\end{equation}
It indicates that the central point of the 0-th row of $\mathbf{H}_{i}$ (i.e., $\operatorname{ind}_{i}$) is determined by the delay-Doppler profile ($l_{i}, \alpha_{i}$) of the $i$-th path, $\alpha_{\max }$, and $k_{\nu}$ jointly. Fig. \ref{sci4-3}(b) shows the central points of all the possible paths with integer delay-Doppler profiles $l_{i}\in[0, l_{\max}]$, $\alpha_{i}\in[-\alpha_{\max }, \alpha_{\max}]$. It can be observed that they do not overlap with each other \cite{bb6}. We further distinguish them with different colors according to their delay shifts, and hence we have $l_{\max}+1$ \textbf{delay blocks} with a size of $2\alpha_{\max }+1$. Each delay block is surrounded by two protection bands with a size of $k_{\nu}$ in the head and tail of it, respectively. 

Define 
\vspace{-0.2em}
\begin{equation}
	\vspace{-0.2em}
L\triangleq(l_{\max}+1)\big(2(\alpha_{\max }+k_{\nu})+1\big)-1
\label{eq7-26-2}
\end{equation}
where $L+1 \ll N$, i.e., the channels are underspread. Then
according to (\ref{eq7-15-1}), there is a large value non-zero band with a size of $L+1$ in each row and column of $\mathbf{H}_{\text{eff}}$, as an example of $\mathbf{H}_{1,1}$ shown in Fig. \ref{sci4-2}.

\emph{\textbf{Remark 2:}} The indices and values of the entries in any large value non-zero band of the effective channel matrix reflect the delay-Doppler profiles and channel gains of the propagation paths, respectively, which means every large value non-zero band contains all the CSI.

\subsection{EPA channel estimation in MIMO-AFDM}
\begin{figure*}[htbp]
	\vspace{-1em}
	\centering
	\includegraphics[width=0.92\textwidth,height=0.480\textwidth]{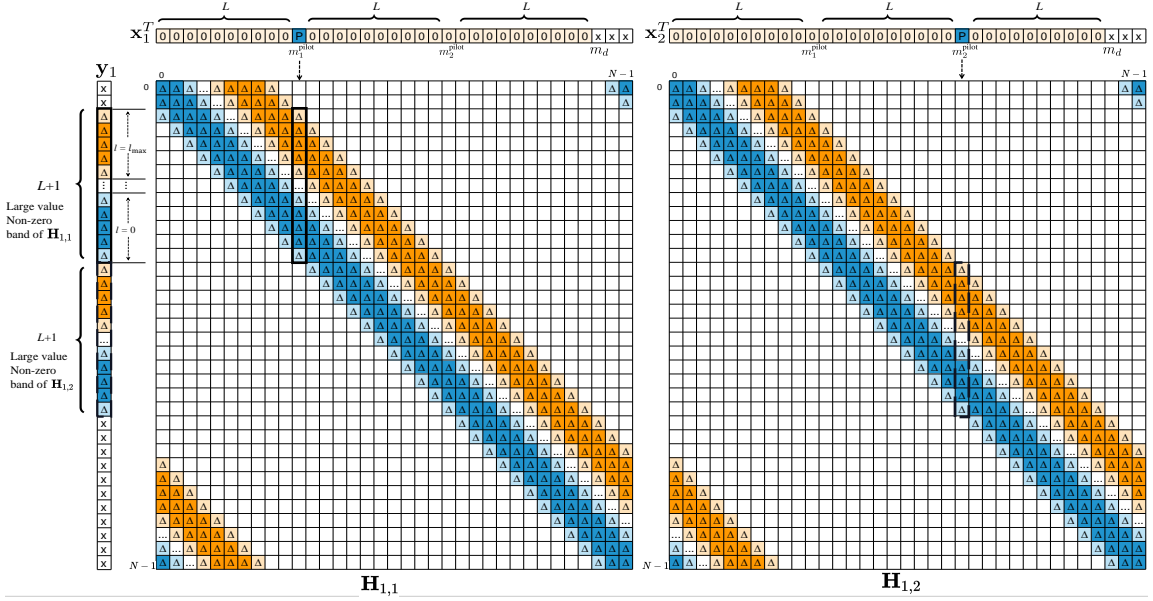}
	\vspace{-1.5em}
	\caption{EPA channel estimation for $2 \times 1$ MISO-AFDM system (‘P’:pilot, ‘0’:guard, ‘x’:data, ‘$\Delta$’: large  non-zero value, ‘blank’ in $\mathbf{H}_{r,t}$: small non-zero value).}
	\label{sci4-2}
	\vspace{-1.8em}
\end{figure*}

We arrange the pilot, guard, and data symbols in the DAFT domain at the $t$-th TA ($t=1,\cdots,N_{t}$) as
\begin{equation}
	x_{t}[m]=\left\{\begin{array}{ll}
		x_{t}^{\text{pilot}},& m = (L+1)t-1\\
		0, & m \in [0,(L+1)N_{t}+L-1] \  \\
		& \quad \ \text{and} \ m \neq (L+1)t-1\\
		x_{t}^{\text{data}}[m], & m\in[m_{d},N-1] \\
	\end{array}\right.
	\label{eq4-43-6}
\end{equation}
where  $x_{t}^{\text{pilot}}$ is the pilot symbol of $t$-th TA with index $m^{\text{pilot}}_{t} = (L+1)t-1$, $x_{t}^{\text{data}}[m]$ is the data symbol in the $m$-th slot of the $t$-th TA, and the index of the first data symbol is $m_{d}=(L+1)N_{t}+L$. An example of  $2 \times 1$ MISO-AFDM system is presented in Fig. \ref{sci4-2}. The main idea of the symbol arrangement in (\ref{eq4-43-6}) is embedding one pilot in the transmitted AFDM symbol at each TA to extract a large value non-zero band from $\mathbf{H}_{r,t}$ at the $r$-th RA. Zero guard symbol bands with a size of $L$ are placed in the adjacency of all the pilots so that the extracted non-zero bands at any RA from different TA can be separated approximately. 

At the $r$-th RA ($r=1,\cdots,N_{r}$), the indices of received symbols $y_{r}[m]$ that are used for channel estimation between the $r$-th RA and the $t$-th TA are 
\begin{equation}
	m\in[\alpha_{\max} +k_{\nu}+ (L+1)(t-1),\alpha_{\max}+k_{\nu}+(L+1)t-1]. 
	\label{eq4-1-22.10.23}
\end{equation}

\subsection{Interference analysis among received pilot symbols}
\label{sec3-3}
We next show the challenges of acquiring CSI from the received pilot symbols. For ease of illustration, we assume in this subsection that the exact or estimated delay-Doppler profile of the $P$ paths is known at the receiver, while their associated channel gains are to be determined. 

We categorize all the propagation paths according to their delay shifts in the following. The subscript $(l,j)$ represents the $j$-th path with delay $l$. Then the  corresponding channel gain, Doppler shift, index indicator, and subchannel matrix can be denoted as $h_{l,j}^{[r,t]}$, $\nu_{l,j}=\alpha_{l,j}+\beta_{l,j}$, $\operatorname{ind}_{l,j}$, and $\mathbf{H}_{(l,j)}$, respectively. Denote the number of paths with delay $l$ as $P_{l} \geq 0$, where $\sum_{l=0}^{l_{\max}}P_{l}=P$. Then (\ref{eq7-15-1}) can be rewritten as
\vspace{-0.2em}
\begin{equation}
	\vspace{-0.5em}
	\mathbf{H}_{r,t} = \sum_{l=0}^{l_{\max}} \sum_{j=1}^{P_{l}} h_{l,j}^{[r,t]} \mathbf{H}_{(l,j)}.
	\label{eq7-16-2}
\end{equation} 

\textbf{Without loss of generality, we attempt to estimate the channel gain of the $j'$-th path with delay $l'$ between the $r$-th RA and $t'$-th TA, i.e., $h_{l',j'}^{[r,t']}$ in the following.} To this end, we consider coordinate $[m,m^{\text{pilot}}_{t'}]$ as the central point in the $m$-th row of  $\mathbf{H}_{(l',j')}$, i.e., $m$ from (\ref{eq4-1-22.10.23}) is chosen elaborately to satisfy
\vspace{-0.5em}
\begin{equation}
	\vspace{-0.5em}
	 m^{\text{pilot}}_{t'}=(m+\operatorname{ind}_{l',j'})_{N}. 
	\label{eq7-13-1}
\end{equation}
Then from (\ref{eq2-111}) and (\ref{eq100}), the received pilot symbol $y_{r}[m]$ can be decomposed into five components apart from the noise $w_{r}[m]$ as
\begin{equation}
	\label{eq4-56}
	\begin{aligned}
		&y_{r}[m] 
	 =
		\sum_{t=1}^{N_{t}}
		\sum_{l=0}^{l_{\max}}
		\sum_{j=1}^{P_{l}}
		\sum_{m'=0}^{N-1}		 
		h_{l,j}^{[r,t]} 
		\mathbf{H}_{(l,j)}[m,m']
		x_{t}[m']+w_{r}[m]=			   
		\underbrace{h_{l',j'}^{[r,t']} 
			\mathbf{H}_{(l',j')}[m,m^{\text{pilot}}_{t'}]
			x_{t'}^{\text{pilot}}}_{\textbf{Desired component}}\\
		&+
		\underbrace{\sum_{j=1, \ j \neq j'}^{P_{l'}}   
			h_{l',j}^{[r,t']} 
			\mathbf{H}_{(l',j)}[m,m^{\text{pilot}}_{t'}]
			x_{t'}^{\text{pilot}}}_{\textbf{IDoI}} +
		\underbrace{
			\sum_{l=0,\ l \neq l'}^{l_{\max}}
			\sum_{j=1}^{P_{l}}  		 
			h_{l,j}^{[r,t']} 
			\mathbf{H}_{(l,j)}[m,m^{\text{pilot}}_{t'}]
			x_{t'}^{\text{pilot}}}_{   \textbf{IDI}} \\
		&+ 
		\underbrace{
			\sum_{t=1, \ t\neq t' }^{N_{t}} 
			\sum_{l=0}^{l_{\max}}
			\sum_{j=1}^{P_{l}}    
			h_{l,j}^{[r,t]} 
			\mathbf{H}_{(l,j)}[m,m^{\text{pilot}}_{t}]	
			x_{t}^{\text{pilot}}}_{\textbf{IPI}} +
		\underbrace{
			\sum_{t=1 }^{N_{t}}
			\sum_{l=0}^{l_{\max}}
			\sum_{j=1}^{P_{l}}
			\sum_{m'=m_{d}}^{N-1}      
			h_{l,j}^{[r,t]}
			\mathbf{H}_{(l,j)}[m,m']	  
			x_{t}^{\text{data}}[m']}_{\textbf{IPDI}}+w_{r}[m]. \\		
	\end{aligned}
\end{equation}

From (\ref{eq7-5-1}) and (\ref{eq7-13-1}) we know that the term with $\mathbf{H}_{(l',j')}[m,m^{\text{pilot}}_{t'}]$ contains the strongest energy of $h_{l',j'}^{[r,t']}$. If we could separate the \textbf{Desired component} from $y_{r}[m]$, $h_{l',j'}^{[r,t']}$ can be recovered given that $m^{\text{pilot}}_{t'}$ and $\mathbf{H}_{(l',j')}[m,m^{\text{pilot}}_{t'}]$ are known at the receiver ($\mathbf{H}_{(l',j')}[m,m^{\text{pilot}}_{t'}]$ can be calculated with (\ref{eq2-13-2})). However, this separation hypothesis is impossible to achieve without being aware of the channel gains of the other paths. Therefore, the residuals of $y_{r}[m]$ should be seen
as interference to the estimation of  $h_{l',j'}^{[r,t']}$, which can be divided into the following four parts: 

\textbf{1)  Inter-Doppler interference (IDoI):} components conveyed by the pilot $x_{t'}^{\text{pilot}}$ and belong to the  
paths with the same delay shift $l'$ but different Doppler shifts (occur in the thrid case of Table \ref{tabel7-31-1}). In this case, according to (\ref{eq7-3-2}), the distance between the central points of $\mathbf{H}_{(l',j')}$ and any $\mathbf{H}_{(l',j)}$ with $j \neq j'$, i.e.,
\begin{equation}
 \delta^{(l',j')}_{(l',j)} \triangleq |\operatorname{ind}_{l',j'}-\operatorname{ind}_{l',j}|_{N}
 \label{eq7-10-1}
\end{equation}
is relatively small, which is within $[1,2\alpha_{\max}]$. Therefore, the magnitude of $\mathbf{H}_{(l',j)}[m,m^{\text{pilot}}_{t'}]$ is comparable to  $\mathbf{H}_{(l',j')}[m,m^{\text{pilot}}_{t'}]$ from the desired component. Furthermore, the energy of pilot symbol is large in practice, making the IDoI term extremely serious. We visualize IDoI on the right side of Fig. \ref{sci4-3}(a) with an example that $l'=l_{\max}$, $j'=1$ ($\alpha_{l_{\max},1}=-1$), $j=2$ ($\alpha_{l_{\max},2}=0$), and $\delta^{(l_{\max},1)}_{(l_{\max},2)}=1$. we can observe that when the two paths have the same delay shift, i.e., their index indicators belong to the same delay block, their subchannel matrice overlap severely.

\textbf{2)  Inter-delay interference (IDI):} components conveyed by the pilot $x_{t'}^{\text{pilot}}$ and belong to the paths with delay shifts $l \neq l'$ (occur in the first and second cases of Table \ref{tabel7-31-1}). The distance between the central points of $\mathbf{H}_{(l',j')}$ and any $\mathbf{H}_{(l,j)}$ with $l \neq l'$, namely $\delta^{(l',j')}_{(l,j)}$, has an lower bound of $2k_{\nu}+1$. Therefore, when the spacing factor $k_{\nu}$ is sufficiently large, the magnitude of $\mathbf{H}_{(l,j)}[m,m^{\text{pilot}}_{t'}]$ is relatively small, which means IDI is much more moderate than IDoI. We visualize IDI on the left side of Fig. \ref{sci4-3}(a) with an example that $l'=0$ ($j'=1$, $\alpha_{0,1}=\alpha_{\max }$), $l=1$ ($j=1$, $\alpha_{1,1}=-\alpha_{\max }$) and $\delta^{(0,1)}_{(1,1)}=2k_{\nu}+1$. We can notice that the two subchannel matrice overlap level is much smaller than the IDoI case. Moreover,  increasing $k_{\nu}$ can further lower the overlap level between the two subchannel matrice and hence suppresses the IDI.

\textbf{3)  Inter-pilot interference (IPI):} components contributed by the pilot symbols $x_{t}^{\text{pilot}}$ from the other TA, i.e., $t \neq t'$. The “blank slot" in AFDM effective channel matrix represents a small non-zero value, and the non-zero bands of effective channel matrice from all the TA are extracted simultaneously at the $r$-th RA in MIMO-AFDM system. Therefore, the pilots symbol from the other TA will affect $y_{r}[m]$, which can be observed from Fig. \ref{sci4-2}. Note that IPI becomes comparable to IDI at the connection of two adjacent large value non-zero bands.
Moreover, increasing $k_{\nu}$ can enlarge the distances between $x_{t}^{\text{pilot}}$ ($t\neq t'$) and the central points of the $m$-th row of all the $\mathbf{H}_{(l,j)}$ and thus lower the magnitude of $\mathbf{H}_{(l,j)}[m,m^{\text{pilot}}_{t}]$, i.e., IPI can be alleviated.

\textbf{4)  Inter-pilot-data interference (IPDI):} components contributed by the data symbols. Similar to IPI, data symbols  from all the TA will affect $y_{r}[m]$. However, since the energy of data symbol is much smaller than the pilot symbol in practice, IPDI is less significant compared to IPI. Moreover, it can be mitigated by increasing $k_{\nu}$ for the same reason.

\emph{\textbf{Remark 3:}} In the presence of fractional Doppler shifts, four types of interference occur when estimating the channel gains from the received pilot symbols. Their relative interference degree is given by
	\vspace{-0.3em}
\begin{equation}
		\vspace{-0.5em}
\textbf{Desired component}\approx	\textbf{IDoI} \gg \textbf{IDI} \ \textgreater \ \textbf{IPI} \ \textgreater \ \textbf{IPDI}.
	\label{eq7-12-1}
\end{equation}
Increasing the $k_{\nu}$ is equivalent to enlarging the distance between two consecutive delay blocks, and hence can only suppress the IPI, IPDI, and IDI, not the most serious IDoI. 

Note that the above attempt to estimate $h_{l',j'}^{[r,t']}$ is carried out under the premise that we have the exact or estimated delay-Doppler profile of the $P$ paths, which is another difficulty that needs to be addressed in advance. The main purpose of this section is to elaborate that estimating the three channel parameters of each propagation path, i.e., the channel gain, the delay shift, and the Doppler shift between all pairs of RA and TA in MIMO-AFDM system with the serious IDoI is extremely intractable.
This motivates us to seek for another way to estimate $\mathbf{H}_{\text {MIMO}}$.

\section{EPA-DR Channel Estimation for MIMO-AFDM}
	\vspace{-0.3em}
\label{sec4}
In this section, we study the diagonal reconstructability of AFDM subchannel channel matrix. Based on that, we propose a low-complexity method named embedded pilot-aided  diagonal reconstruction (EPA-DR) to estimate $\mathbf{H}_{\text {MIMO}}$. 

\subsection{Diagonal reconstructability of AFDM subchannel matrix}

We first investigate the diagonal reconstructability of subchannel matrix $\mathbf{H}_{(l,j)}$, $\forall \ l\in[0,l_{\max}]$ and $j\in[0,P_{l}]$. Lets define the \textbf{transform factor} between $\mathbf{H}_{(l,j)}[m,m']$ and $\mathbf{H}_{(l,j)}[(m+1)_{N}, (m'+1)_{N}]$ as
\begin{equation}
\mathcal{T}(l,\nu_{l,j},m, m') = \frac{\mathbf{H}_{(l,j)}[(m+1)_{N}, (m'+1)_{N}]}{\mathbf{H}_{(l,j)}[m,m']}.
	\label{eq7-13-3} 
\end{equation}
Substituting (\ref{eq2-13-2}) into (\ref{eq7-13-3}), we have
\begin{equation}
	\mathcal{T}(l,\nu_{l,j},m, m') = \frac{\mathcal{C}(l, (m+1)_{N}, (m'+1)_{N})}{\mathcal{C}(l, m, m')}\frac{\mathcal{F}(l,\nu_{l,j}, (m+1)_{N}, (m'+1)_{N})}{\mathcal{F}(l,\nu_{l,j}, m, m')}.
	\label{eq7-13-4} 
\end{equation}
Since
\begin{equation}
	\begin{aligned}
\mathcal{F}(l,\nu_{l,j}, &(m+1)_{N}, (m'+1)_{N})\\ &=\frac{e^{j 2 \pi\left((m+1)_{N}+\operatorname{ind}_{l,j}-(m'+1)_{N}+\beta_{l,j}\right)}-1}{e^{j \frac{2 \pi}{N}\left((m+1)_{N}+\operatorname{ind}_{l,j}-(m'+1)_{N}+\beta_{l,j}\right)}-1}=\frac{e^{j 2 \pi\left(m+\operatorname{ind}_{l,j}-m'+\beta_{l,j}+\lambda(m,m')N \right)}-1}{e^{j \frac{2 \pi}{N}\left(m+\operatorname{ind}_{l,j}-m'+\beta_{l,j}+\lambda(m,m')N\right)}-1}
\end{aligned}
\label{eq7-14-1} 
\end{equation}
where variable $\lambda(m,m')=0,1,2$ is introduced due to the $N$ modulus operation in $(m+1)_{N}$ and $(m'+1)_{N}$, we have
\begin{equation}
		\mathcal{F}(l,\nu_{l,j}, (m+1)_{N}, (m'+1)_{N})=\frac{e^{j 2 \pi\left(m+\operatorname{ind}_{l,j}-m'+\beta_{l,j} \right)}-1}{e^{j \frac{2 \pi}{N}\left(m+\operatorname{ind}_{l,j}-m'+\beta_{l,j}\right)}-1}=\mathcal{F}(l,\nu_{l,j}, m, m').
	\label{eq7-14-2} 
\end{equation}
Substituting (\ref{eq7-14-2}) and (\ref{eq2022.11.09}) into (\ref{eq7-13-4}), we have
\begin{equation}
	\begin{aligned}
		\mathcal{T}(l,\nu_{l,j},m, m') &= \frac{\mathcal{C}\big(l, (m+1)_{N}, (m'+1)_{N}\big)}{\mathcal{C}(l, m, m')}	= e^{j \frac{2 \pi}{N}\big(( m'-(m'+1)_{N})l 
			+Nc_{2}\big[((m'+1)_{N})^{2}+m^{2}-((m+1)_{N})^{2}-m'^{2}\big]\big)} \\
		&=\left\{\begin{array}{ll}
			e^{j \frac{2 \pi}{N}\big(-l+2Nc_{2}(m'-m)\big)},& m \textless N-1, m'\textless N-1\\
			
			e^{j \frac{2 \pi}{N}\big(-l+Nc_{2}(m^{2}+2m'+1)\big)}, & m = N-1, m'\textless N-1   \\
			
			e^{j \frac{2 \pi}{N}\big(-l-Nc_{2}(m'^{2}+2m+1)\big)}, & m \textless N-1, m'= N-1  \\
			
			e^{-j \frac{2 \pi}{N}l}, & m = N-1, m'= N-1. \\
		\end{array}\right.
	\end{aligned}
	\label{eq4-38-6} 
\end{equation}
It is important to notice that $\mathcal{T}(l,\nu_{l,j}, m, m')$ has no relevance to the Doppler shift $\nu_{l,j}$, \textbf{which means $\mathcal{T}(l,\nu_{l,j}, m, m')$ can be represented as $\mathcal{T}(l, m, m')$ instead}. This makes a fundamental contribution to the robustness of the proposed channel estimation scheme to IDoI and will be illustrated in the next subsection.  

Consequently, according to (\ref{eq7-13-3}), as long as $\mathbf{H}_{(l,j)}[m,m']$ is obtained, $\mathbf{H}_{(l,j)}[(m+1)_{N}, (m'+1)_{N}]$ can be calculated directly with $\mathcal{T}(l,m, m')$ given in (\ref{eq4-38-6}) as
\begin{equation}
	\mathbf{H}_{(l,j)}[(m+1)_{N}, (m'+1)_{N}] = \mathcal{T}(l, m, m') \mathbf{H}_{(l,j)}[m,m'].
	\label{eq7-16-1} 
\end{equation}
This implies that as long as we acquire an arbitrary column of $\mathbf{H}_{(l,j)}$, the whole $\mathbf{H}_{(l,j)}$ can be reconstructed “diagonally”. We refer to this unique characteristic of $\mathbf{H}_{(l,j)}$ as \textbf{diagonal reconstructability}.

\subsection{EPA-DR channel estimation scheme}
We next explore the diagonal reconstructability to  estimate effective channel matrix $\mathbf{H}_{r,t'}$ with the received pilot symbols $y_{r}[m]$ given in (\ref{eq4-1-22.10.23}). 

Following the consideration in Section \ref{sec3-3} that
coordinate $[m,m^{\text{pilot}}_{t'}]$ is the central point in the $m$-th row of $\mathbf{H}_{(l',j')}$, an estimation of $\mathbf{H}_{r,t'}[m,m^{\text{pilot}}_{t'}]$ can be obtained with (\ref{eq4-56}) as
\vspace{-0.35em}
\begin{equation}
\begin{aligned}
&\mathbf{\hat{H}}_{r,t'}[m,m^{\text{pilot}}_{t'}]=
\frac{y_{r}[m]}{x_{t'}^{\text{pilot}}}\\
& =
\underbrace{\underbrace{\sum_{j=1}^{P_{l'}} h_{l',j}^{[r,t']} \mathbf{H}_{(l',j)}[m,m^{\text{pilot}}_{t'}]}_{\text{Main component of} \ \mathbf{H}_{r,t'}[m,m^{\text{pilot}}_{t'}]}+
	\underbrace{\sum_{l=0,l\neq l'}^{l_{\max}} \sum_{j=1}^{P_{l}} h_{l,j}^{[r,t']} \mathbf{H}_{(l,j)}[m,m^{\text{pilot}}_{t'}]}_{\text{Normalized \textbf{IDI} from (\ref{eq4-56})}}}_{\mathbf{H}_{r,t'}[m,m^{\text{pilot}}_{t'}]}+\frac{\text{IPI}+\text{IPDI}+w_{r}[m]}{x_{t'}^{\text{pilot}}}
\end{aligned}
\label{eq7-17-1}
\vspace{-0.15em}
\end{equation} 
where the first term is composed of the desired component and the IDoI from (\ref{eq4-56}), while the second term comes from the IDI. They constitute the real $\mathbf{H}_{r,t'}[m,m^{\text{pilot}}_{t'}]$ and the first term contributes the most energy of $\mathbf{\hat{H}}_{r,t'}[m,m^{\text{pilot}}_{t'}]$ since $(m,m^{\text{pilot}}_{t'})$ belongs to the $l'$-th delay block. \textbf{Therefore, the original strong IDoI term when estimating $h_{l',j'}^{[r,t']}$  should no longer be considered as interference.} The actual interference for estimating $\mathbf{H}_{r,t'}[m,m^{\text{pilot}}_{t'}]$ are the third term in (\ref{eq7-17-1}).

With $\mathbf{\hat{H}}_{r,t'}[m,m^{\text{pilot}}_{t'}]$ in hand, an estimation of  $\mathbf{H}_{r,t'}[(m+1)_{N},(m^{\text{pilot}}_{t'}+1)_{N}]$ can be acquired by exploring (\ref{eq7-16-1}) as 
\begin{equation}
	\mathbf{\hat{H}}_{r,t'}[(m+1)_{N},(m^{\text{pilot}}_{t'}+1)_{N}]=\mathcal{T}(l',m,m^{\text{pilot}}_{t'})\mathbf{\hat{H}}_{r,t'}[m,m^{\text{pilot}}_{t'}]
	\label{eq7-17-3}
\end{equation}
given that 
\begin{equation}
	\mathcal{T}(l',m,m^{\text{pilot}}_{t'})\underbrace{\sum_{j=1}^{P_{l'}} h_{l',j}^{[r,t']} \mathbf{H}_{(l',j)}[m,m^{\text{pilot}}_{t'}]}_{\text{Main component of} \ \mathbf{H}_{r,t'}[m,m^{\text{pilot}}_{t'}]}\\
	=\sum_{j=1}^{P_{l'}} h_{l',j}^{[r,t']} \mathbf{H}_{(l',j)}[(m+1)_{N},(m^{\text{pilot}}_{t'}+1)_{N}]
\label{eq7-17-5}
\end{equation}
is the main component of  $\mathbf{H}_{r,t'}[(m+1)_{N},(m^{\text{pilot}}_{t'}+1)_{N}]$. This because coordinate $[(m+1)_{N},(m^{\text{pilot}}_{t'}+1)_{N}]$ still belongs to the $l'$-th delay block.

\emph{\textbf{Remark 4:}} The residual errors of $\mathbf{\hat{H}}_{r,t'}[(m+1)_{N},(m^{\text{pilot}}_{t'}+1)_{N}]$ in (\ref{eq7-17-3}) are the IDI, IPI, IPDI, and noise $w_{r}[m]$, which are weighted by $\mathcal{T}(l',m,m^{\text{pilot}}_{t'})$ and inherited from $\mathbf{\hat{H}}_{r,t'}[m,m^{\text{pilot}}_{t'}]$. According to Remark 3, their energy is much smaller than the main component and hence can be modeled as additive noise. It is worth emphasizing that there is no IDoI anymore, which means the IDoI can be avoided naturally.

Consequently, by performing (\ref{eq7-17-3}) iteratively, i.e.,
\begin{equation}
\begin{aligned}
	&\mathbf{\hat{H}}_{r,t'}[(m+i+1)_{N},(m^{\text{pilot}}_{t'}+i+1)_{N}]\\
	&=\mathcal{T}(l',(m+i)_{N},(m^{\text{pilot}}_{t'}+i)_{N})  \mathbf{\hat{H}}_{r,t'}[(m+i)_{N},(m^{\text{pilot}}_{t'}+i)_{N}], \
	 i = 0,\cdots,N-2
\end{aligned}
\label{eq7-26-1}
\end{equation}
$\mathbf{\hat{H}}_{r,t'}[(m+1)_{N},(m^{\text{pilot}}_{t'}+1)_{N}]$, $\cdots$, $\mathbf{\hat{H}}_{r,t'}[(m+N-1)_{N},(m^{\text{pilot}}_{t'}+N-1)_{N}]$ can be obtained successively, where the main component of $\mathbf{H}_{r,t'}[(m+1)_{N},(m^{\text{pilot}}_{t'}+1)_{N}]$, $\cdots$, $\mathbf{H}_{r,t'}[(m+N-1)_{N},(m^{\text{pilot}}_{t'}+N-1)_{N}]$ are reserved therein. Moreover, since all the transform factors provided in (\ref{eq4-38-6}) are complex exponentials with energy one, the inherited errors of  $\mathbf{\hat{H}}_{r,t'}[(m+1)_{N},(m^{\text{pilot}}_{t'}+1)_{N}]$, $\cdots$, $\mathbf{\hat{H}}_{r,t'}[(m+N-1)_{N},(m^{\text{pilot}}_{t'}+N-1)_{N}]$ have the same magnitude, i.e., no error accumulation occurs during the iterative estimation process.

In addition, to combat the interference of the noise $w_{r}[m]$ in $y_{r}[m]$, we should first conduct the threshold-based magnitude detection, which is given by
\begin{equation}
\mathbf{\hat{H}}_{r,t}[m,m^{\text{pilot}}_{t}] = \left\{\begin{array}{ll}
	\frac{y_{r}[m]}{x_{t}^{\text{pilot}}},& |y_{r}[m]| \geq \zeta \\
			0, & otherwise. \\ 
\end{array}\right.
\label{eq4-37}
\end{equation}
The influence of the threshold $\zeta$ selection on the channel estimation accuracy will be shown in Section \ref{sec4-3}. By performing the above diagonal estimation on all the received pilot symbols $y_{r}[m]$ with magnitude exceeds the threshold, the estimation of large value non-zero bands in all the columns of $\mathbf{H}_{r,t'}$ can be obtained. We name the above procedures used to estimate the effective channel matrix $\mathbf{H}_{r,t'}$ as embedded pilot-aided diagonal reconstruction (EPA-DR) scheme and summarize it in Algorithm \ref{algorithm-1}.
\begin{algorithm}
	\caption{\textit{EPA-DR} scheme for $\mathbf{H}_{\text{MIMO}}$ estimation}
	\label{algorithm-1}
	
	\textbf{-Transmitter:}\\
	\KwIn{$l_{\max}$, $\alpha_{\max}$, $c_{1}$, $c_{2}$, $N_{t}$, and $m^{\text{pilot}}_{t}$, $t= 1,\cdots,N_{t}$.}  
	
	\For{$t = 1,2,\cdots, N_{t}$}{Arrange the pilot, guard, and data symbols at the $t$-th transmit antenna with (\ref{eq4-43-6}).} 

	\textbf{-Receiver:}\\ 
	\LinesNumbered  
	\KwIn{$l_{\max}$, $\alpha_{\max}$, $c_{1}$, $c_{2}$,  $N_{t}$, $N_{r}$, $\zeta$, and $m^{\text{pilot}}_{t}$, $t= 1,\cdots,N_{t}$.}
	\For{$r = 1,2,\cdots, N_{r}$}{
		\For{$t = 1,2,\cdots, N_{t}$}{
			\textbf{Step 1:}  Obtain the large value  non-zero band in the $m^{\text{pilot}}_{t}$-th column of $\mathbf{H}_{r,t}$ from  $\mathbf{y}_{r}$ according to (\ref{eq4-1-22.10.23}); \\
			
			\textbf{Step 2:} Perform threshold-based magnitude detection using (\ref{eq4-37});\\
			
			\textbf{Step 3:} Reconstruct  $\mathbf{\hat{H}}_{r,t}$ iteratively using (\ref{eq7-26-1}).\\
		}
	}
	\KwOut{Assemble $\mathbf{\hat{H}}_{r,t}$ into $\mathbf{\hat{H}}_{\text{MIMO}}$ using (\ref{eq.77}).}
\end{algorithm}

With the above EPA-DR scheme, we can estimate the effective channel matrices between all pairs of receive antennas and transmit antennas at the receiver simultaneously. The estimated $\mathbf{\hat{H}}_{\text {MIMO}}$ are then used to detect the received data symbols $y_{r}[m]$, $m \notin [\alpha_{\max} +k_{\nu},\alpha_{\max}+k_{\nu}+(L+1)tN_{t}-1]$, which is given by
\begin{equation}
	\label{eq4-57-10}
		y_{r}[m]= 	 		
		\sum_{t=1 }^{N_{t}}
		\sum_{m'=\tilde{m}_{d} }^{N-1}  
		\sum_{l=0}^{l_{\max}}
		\sum_{j=1}^{P_{l'}}    
		h_{l,j}^{[r,t]}
		\mathbf{H}_{(l,j)}[m,m'] 
		x_{t}^{\text{data}}[m'] +
		\underbrace{
			\sum_{t=1}^{N_{t}} 
			\sum_{l=0}^{l_{\max}}
			\sum_{j=1}^{P_{l}}  		 
			h_{l,j}^{[r,t]} 
			\mathbf{H}_{(l,j)}[m,m^{\text{pilot}}_{t}]
			x_{t}^{\text{pilot}}}_{  \textbf{IPDI}} +w_{r}[m].	
\end{equation}
We can observe from (\ref{eq4-57-10}) that, due to the fractional Doppler, the received data symbols are interfered by all the pilots,  which can also be alleviated by enlarging the spacing factor $k_{\nu}$.

\subsection{Performance assessment of EPA-DR algorithm}

\emph{\textbf{1)  Overhead analysis}}: The pilot and guard overhead for each transmit antenna in an $N_{t} \times N_{r}$ MIMO-AFDM system is given by
\begin{equation}
	O_{\text{MIMO-AFDM}} =(N_{t}+1)L+N_{t}= (N_{t}+1)(l_{\max}+1)\big(2(\alpha_{\max }+k_{\nu})+1\big)-1.
	\label{eq4-101}
\end{equation}
While the $N_{t} \times N_{r}$ MIMO-OTFS counterpart is \cite{b4}
\begin{equation}
O_{\text{MIMO-OTFS}}=\big((N_{t}+1)l_{\max}+N_{t}\big)\big(4(\alpha_{\max }+k_{\nu})+1\big)
\label{eq7-28-1}
\end{equation}
which is nearly twice of $O_{\text{MIMO-AFDM}}$.
This implies that AFDM maintains its advantage over OTFS on less channel estimation overhead in MIMO system.

\emph{\textbf{2)  Computational complexity analysis}}: The total computation complexity of the EPA-DR scheme is dominated by \textbf{Step 3} at the receiver, which requires at most only $(L+1)(N-1)$ complex multiplications (assume all the magnitude of received pilot symbols exceed the threshold).

It is worth emphasizing that once the AFDM parameters $c_{1}$ and $c_{2}$ are determined, all the transform factors needed in \textbf{Step 3} can be calculated with (\ref{eq4-38-6}) only once in advance at the receiver despite the ever-changing channels ($l_{i}$ can be obtained from the bijective relationship between the coordinate ($m$, $m'$) and the  $l_{i}$-th delay block). To be specific, when $m \ \textless \ N-1$ and $m'\ \textless \ N-1$,  $\mathcal{T}(l_{i},m, m')$ is determined by $m-m'$, i.e., only four transform factors are required to be calculated for each received pilot symbol (for $m=m'$, it is one). Therefore, the total number of transform factors needed to be calculated in practice is $4L+1$, whose computation complexity is ignorable.

\emph{\textbf{3)  Comparison between EPA-DR and EPA-AML \cite{bb6} schemes}}: In EPA-AML scheme, the delay and Doppler shifts of the $P$ paths are first estimated with brute force search, and then the associate channel gains are estimated. Finally, the AFDM effective channel matrix is calculated with the three estimated parameters of each path according to (\ref{eq7-15-1}). Therefore, the number of paths should be known in advance, and high computation complexity is unavoidable in the brute force search and effective channel matrix calculation. More importantly, it suffers from serious IDoI as we unveiled in Section \ref{sec3-3}.

In contrast, the proposed EPA-DR scheme does not need to know the number of paths in advance since a threshold-based magnitude detection procedure is adopted. Moreover, the AFDM effective channel matrix is obtained directly without estimating the three channel sparameters and hence lowers the computation complexity significantly. Furthermore, the serious IDoI can be avoided inherently by exploring the diagonal reconstructability of AFDM subchannel matrix. These outstanding advantages of EPA-DR over EPA-AML ensure its feasibility, especially in multiple-antenna-based AFDM systems.

\section{Orthogonal Resource Allocation and Channel Estimation for AFDMA}
\label{sec5}
Inspired by the symbol arrangement in EPA-DR channel estimation scheme, we propose a practical and efficient resource allocation scheme for AFDMA system. For ease of illustration, we consider $N_{U}$ users with single antenna, while the base station (BS) has $N_{\text{BS}}$ antennas. 

Different from the point-to-point MIMO-AFDM system discussed above, the delay-Doppler profiles between the BS and different users are independent of each other. Let $\alpha^{[\text{BS},u]}_{\max }$ and $l^{[\text{BS},u]}_{\max}$ denote the maximum Doppler and maximum delay between the BS and the $u$-th user, respectively, $\tilde{\alpha}_{\max}=max\{\alpha^{[\text{BS},u]}_{\max }| u\in[1, N_{u}]\}$. Then the AFDM parameter $c_{1}$ should be set as $\frac{2(\tilde{\alpha}_{\max}+k_{\nu})+1}{2N}$. Define $L_{u} \triangleq (l^{[\text{BS},u]}_{\max} +1)\big(2(\tilde{\alpha}_{\max}+k_{\nu})+1\big)-1$, where  $L_{u}$ satisfy
\begin{equation}
	L_{1} \leq L_{2} \leq \cdots \leq L_{N_{u}} \leq L_{\max}
	\label{eq4-4-1-11}
\end{equation}
i.e., the users are sorted in order of  value $L_{u}$. The parameter $L_{\max}$ is an arbitrary integer that satisfies equation (\ref{eq4-4-1-11}). Suppose that the BS is aware of $L_{\max}$ and all the $L_{u}$, while each user only knows $L_{\max}$ and its own $L_{u}$.

\vspace{-1em}
\subsection{Downlink}
\begin{figure}[h]
	\begin{minipage}[t]{0.5\textwidth}
		\centering
		\includegraphics[scale=0.29]{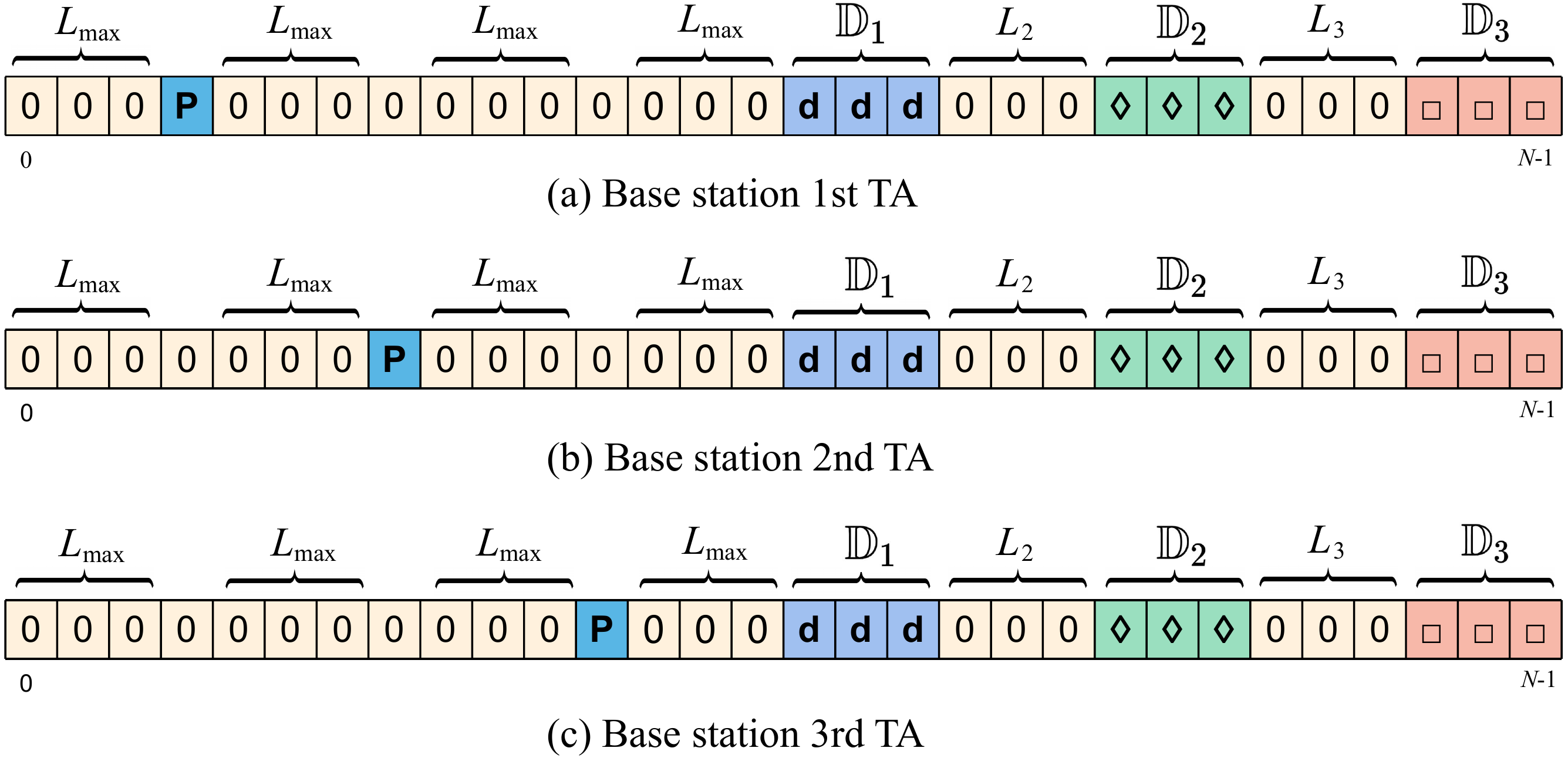}
		\vspace{-1.5em}
		\caption{AFDMA downlink system with three single-antenna users and a three-antenna BS (‘P’: pilot, ‘0’: guard, ‘d’: data symbol for the user 1, ‘$\diamond$’: data symbol for the user 2, ‘$\square$’: data symbol for the user 3).}
		\label{4-4-1}
	\end{minipage}
	\quad
	\begin{minipage}[t]{0.45\textwidth}
		\centering
		\includegraphics[scale=0.29]{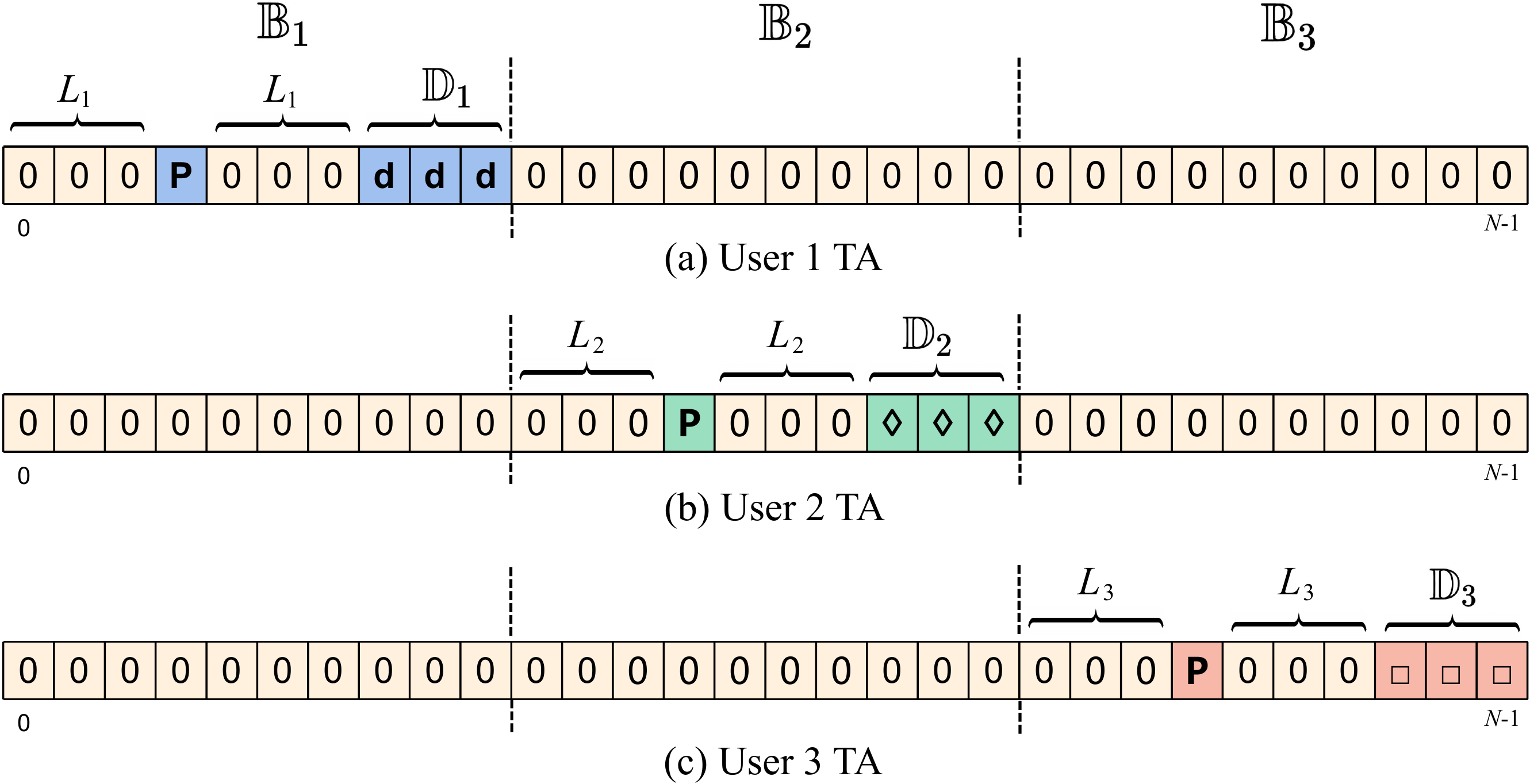}
		\vspace{-3.25em}
		\caption{AFDMA uplink system with three single-antenna users and a three-antenna BS (‘P’: pilot, ‘0’: guard, ‘d’: data symbol of the user 1, 
			‘$\diamond$’: data symbol of the  user 2, ‘$\square$’: data symbol of the user 3).}
		\label{4-4-2}
	\end{minipage}
\vspace{-1.5em}
\end{figure}

In downlink communication, the symbol arrangement at the $b$-th transmit antenna ($b=1,\cdots, N_{\text{BS}}$) in the BS is 
\begin{equation}
	x_{b}[m]=\left\{\begin{array}{ll}
		x_{b}^{\text{pilot}},& m = (L_{\max}+1)b-1\\
		x_{b,u}^{\text{data}}[m], & m\in 
		\mathbb{D}_{u}, u=1,\cdots, N_{u}\\
		0, & otherwise \\
	\end{array}\right.
	\label{eq4-4-1}
\end{equation}
where $x_{b}^{\text{pilot}}$ is the pilot, $\mathbb{D}_{u}$ is the resource block allocated to the $u$-th user for data transmission (each user is assumed to know their own $\mathbb{D}_{u}$), $x_{b,u}^{\text{data}}[m]$ are the corresponding data symbols. Guard symbol blocks (GSB) are inserted among the resource blocks to avoid  inter-user interference (IUI). Each pilot should be surrounded by two GSBs with a size no less than $L_{N_{u}}$, while the $u$-th resource block $\mathbb{D}_{u}$ should be protected by two GSBs with a size no less than $L_{u}$. When these conditions are satisfied, all the users can separate their own received data symbols from the others, making the data detection for each user simpler. 
An example of AFDMA downlink system with three single-antenna users and a three-antenna BS is presented in Fig. \ref{4-4-1}\footnote{Although the data symbols for user 1 and user 2 will coincide with each other at user 3, it is acceptable since the target data symbols of user 3 are not contaminated. Moreover, if we shift the orders of $\mathbb{D}_{2}$ and $\mathbb{D}_{3}$ in Fig. \ref{4-4-1}, then the GSB with the size of $L_{2}$ need to be enlarged to $L_{3}$, i.e., fewer data symbols can be transmitted.}.

The pilot and guard overhead for one TA of BS in downlink communication is given by
\begin{equation}
	O_{\text{downlink}} =(N_{\text{BS}}+1)L_{\max}+N_{\text{BS}}+\sum_{u=2}^{N_{u}}L_{u}.
	\label{eq4-4-2}
\end{equation}
The size of $\mathbb{D}_{u}$ ($u=1,\cdots,N_{u}$) can be tuned flexibly according to the throughput demands of different users, as long as their summation satisfies
\begin{equation}
	\sum_{u=1}^{N_{u}}|\mathbb{D}_{u}|=N-O_{\text{downlink}}.
\end{equation}
Based on the  symbol arrangement in (\ref{eq4-4-1}), each user exploits the associated received symbols
for EPA-DR channel estimation and data detection.

\vspace{-0.5em}
\subsection{Uplink}
We first divide the entire AFDM frame into $N_{u}$ orthogonal resource  blocks  as  $\mathbb{B}_{1},  \mathbb{B}_{2}, \cdots,\mathbb{B}_{N_{u}}$ according to the throughput demands of different users (suppose that the BS is aware of the resource blocks allocation thoroughly, while each user only knows its own resource block $\mathbb{B}_{u}$), where $\sum_{u=1}^{N_{u}}|\mathbb{B}_{N_{u}}|=N$. Then each user arranges its pilot and data block $\mathbb{D}_{u}$ in its own resource block. Similar to the downlink case, GSBs should be placed appropriately to avoid IUI. An example of AFDMA uplink system with three single-antenna users and a three-antenna BS is presented in Fig. \ref{4-4-2}. The pilot and guard overhead of the $u$-th user in uplink communication is  $O^{u}_{\text{uplink}} =2 L_{u}$.

With the  above symbol arrangement, the received symbols corresponding to different users can be separated in the BS perfectly.
The BS explores the associated received symbols to perform EPA-DR channel estimation and data detection for each user. Note that in uplink communication, sorting the users in the order of their $L_{u}$ values is not necessary.

A simpler solution is to replace all the $L_{u}$s with $L_{\max}$ roughly in both downlink and uplink communications, and the users ordering in (\ref{eq4-4-1-11}) is no longer necessary. However, this simplification will incur spectral efficiency degradation. 

The main idea of the above downlink and uplink symbol arrangements is reducing the guard symbols used to ensure orthogonal resource allocation and channel estimation to the greatest extent. It is done by flexibly exploring  the differences in the delay-Doppler profiles between the BS and all the users, i.e., the sizes of the non-zero bands of different effective channel matrices. 
Compared to the orthogonal resource allocation based OTFS-MA system proposed in \cite{bb105}, which suffers from excessive guard symbol overhead due to the 2D structure of OTFS,
the AFDMA system designed above is efficient and easy to implement. Meanwhile, it preserves a large degree of freedom for the BS and users to accommodate various communication requirements. Therefore, we can conclude that AFDMA is a promising solution for multiple access in doubly selective channels.

\vspace{-0.8em}
\section{Simulation Results}
\label{secResults}
In this section, following the vectorized input-output relationship in equation (\ref{eq.8}), we present the performance of MIMO-AFDM in terms of BER with ideal and our estimated CSI. Both integer  and fractional Doppler cases are considered. Each transmit antenna transmits independent information symbols, and the channel gains of all paths follow the distribution of $\mathcal{C N}(0,1 / P)$.

In the case of large frame size, many low-complexity detectors proposed for AFDM and OTFS can be applied to MIMO-AFDM directly \cite{bb102,bb13,  bb22.10.24.5, bb22.10.29.6}. We adopt the widely-used message passing (MP) detector proposed in \cite{bb13} with a complexity order of $\mathcal{O}(n_{\text{iter}}NN_{r}S|\mathbb{A}|)$, where $n_{\text{iter}}$ is the maximum number of iteration, $S \leq L+1$ is the number of received pilot symbols whose energy exceed the preset threshold, $|\mathbb{A}|$ represents the modulation order.

\vspace{-1.5em}
\subsection{Perfect CSI}
\label{secResults_idealCSI}

\begin{figure}[h]
	\vspace{-1.3em}
	\begin{minipage}[t]{0.5\textwidth}
		\centering
		\includegraphics[scale=0.59]{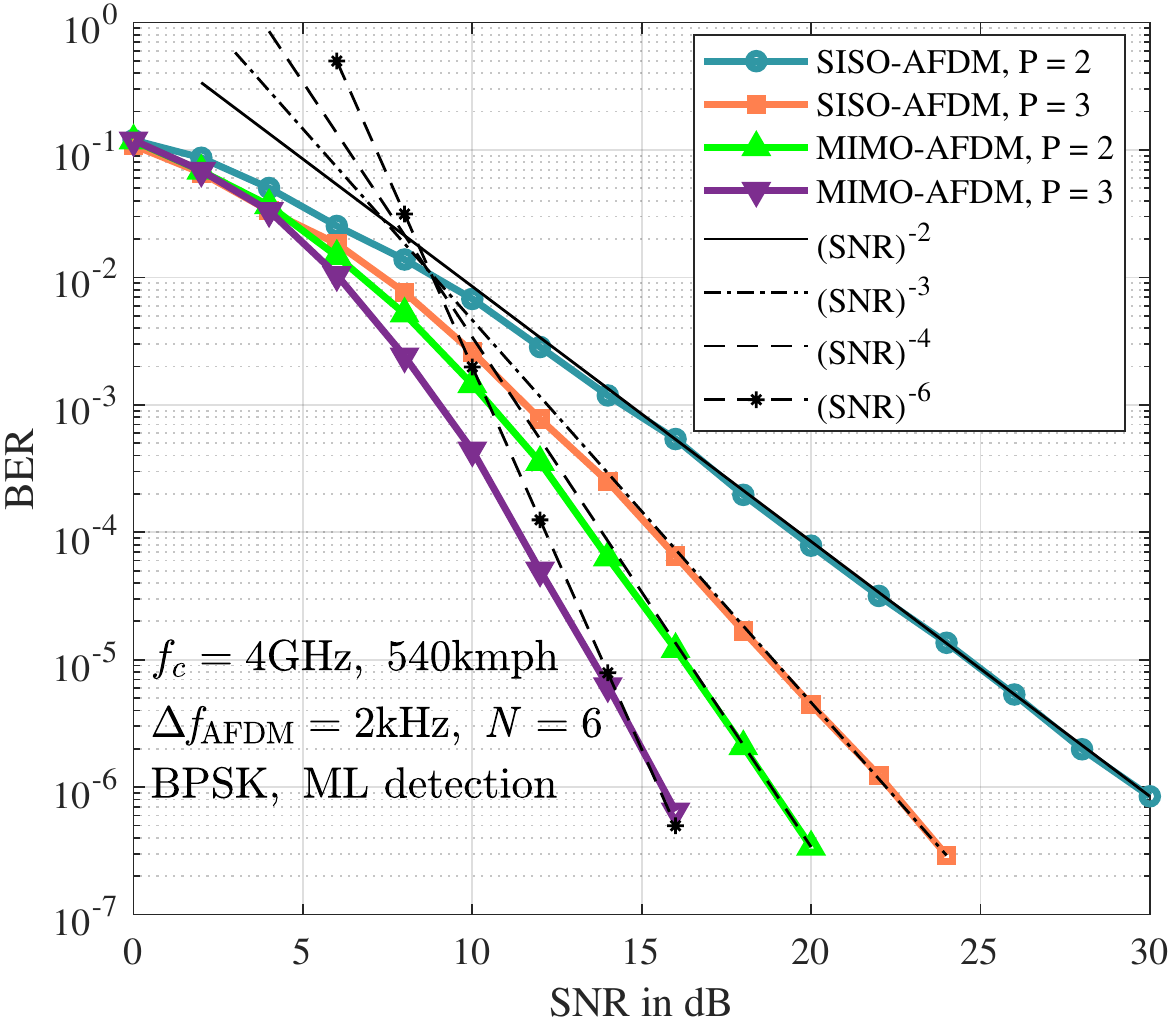}
		\vspace{-1.3em}
		\caption{BER performance of SISO-AFDM and $2 \times 2$ MIMO-AFDM systems with integer Doppler. The normalized delay-Doppler profiles $(l_{i},\nu_{i})$ of two paths: (0, 0) and (1, 1); three paths: (0, 0), (0, 1) and (1, 1).}
		\label{sci3-2}
	\end{minipage}
	\quad
	\begin{minipage}[t]{0.45\textwidth}
		\centering
		\includegraphics[scale=0.5]{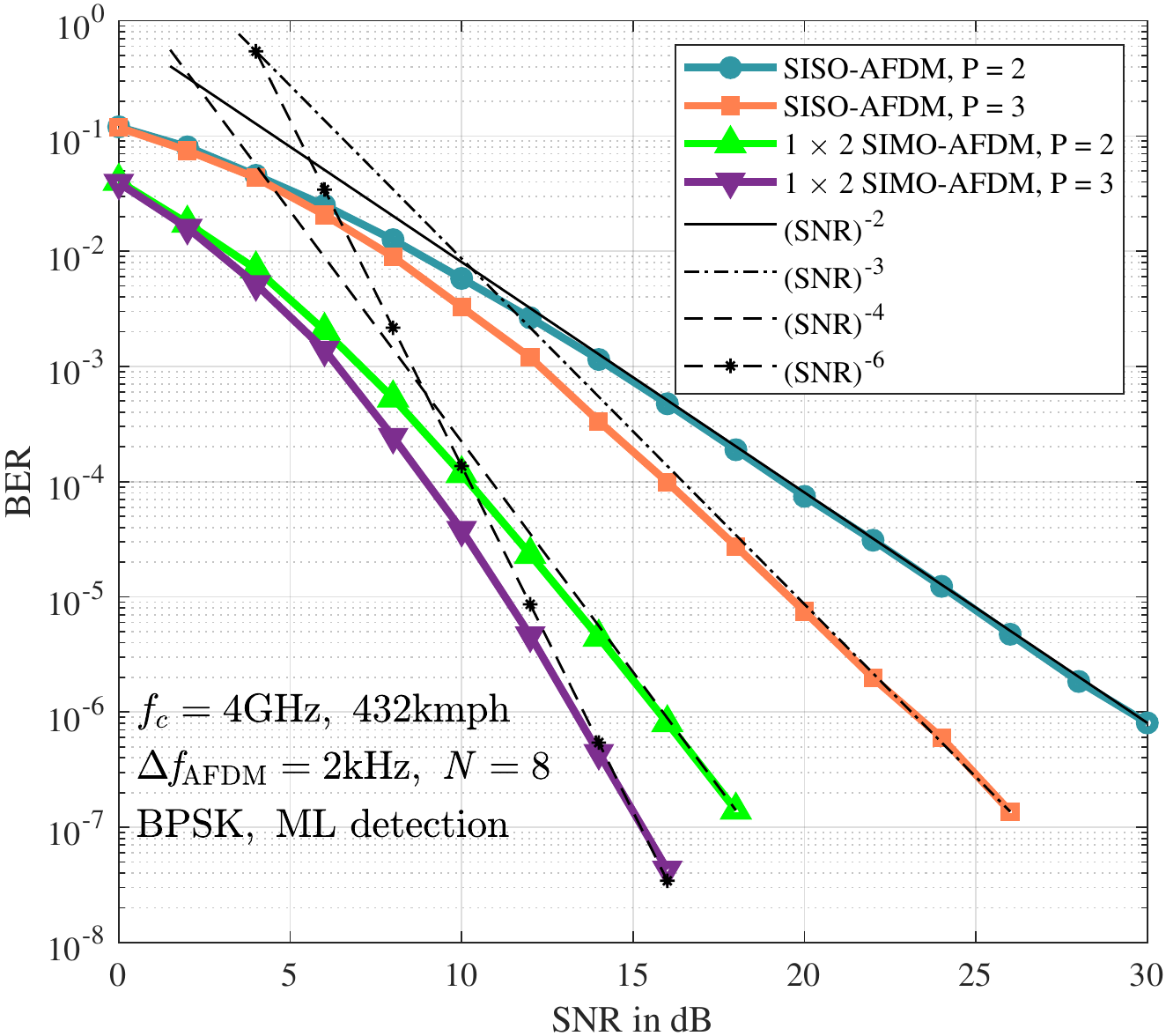}
		\vspace{-1.3em}
		\caption{BER performance of SISO-AFDM and $1 \times 2$ SIMO-AFDM systems with fractional Doppler. The normalized delay-Doppler profiles $(l_{i},\nu_{i})$ of two paths: (0, 0) and (1, 0.5); three paths: (0, 0), (0, 0.5) and (1, 0.8).}
		\label{sci3-4}
	\end{minipage}
	\vspace{-1.3em}
\end{figure}

We first apply maximum likelihood (ML) detector to verify the diversity order of MIMO-AFDM with  small frame sizes. Fig. \ref{sci3-2} shows the BER performance of SISO-AFDM and $2\times2$ MIMO-AFDM systems with integer Doppler ($\beta_{i}= 0$ and $k_{\nu}=0$). Without losing of generality, we consider $N_{\text{AFDM}}=6$, $P=2$, and $3$. Asymptotic lines with slopes of 2 $((\text{SNR})^{-2})$, 3 $((\text{SNR})^{-3})$, 4 $((\text{SNR})^{-4})$, and 6 $((\text{SNR})^{-6})$ are plotted. We can observe that the MIMO-AFDM system outperforms SISO-AFDM system thanks to the space diversity gain from multiple RAs. Meanwhile, the multiple TAs of MIMO-AFDM system bring in the advantages of linear increment in spectral efficiency with the number of TAs. Furthermore, the diversity orders achieved by SISO-AFDM in channels with two and three paths are 2 and 3 respectively, while the  $2 \times 2$ MIMO-AFDM counterparts are 4 and 6 respectively, verifying the validity of Theorem 1. Fig. \ref{sci3-4} shows that Theorem 1 also holds for the fractional Doppler case.

\begin{figure}[h]
	\vspace{-1.3em}
	\begin{minipage}[t]{0.5\textwidth}
		\centering
		\includegraphics[scale=0.56]{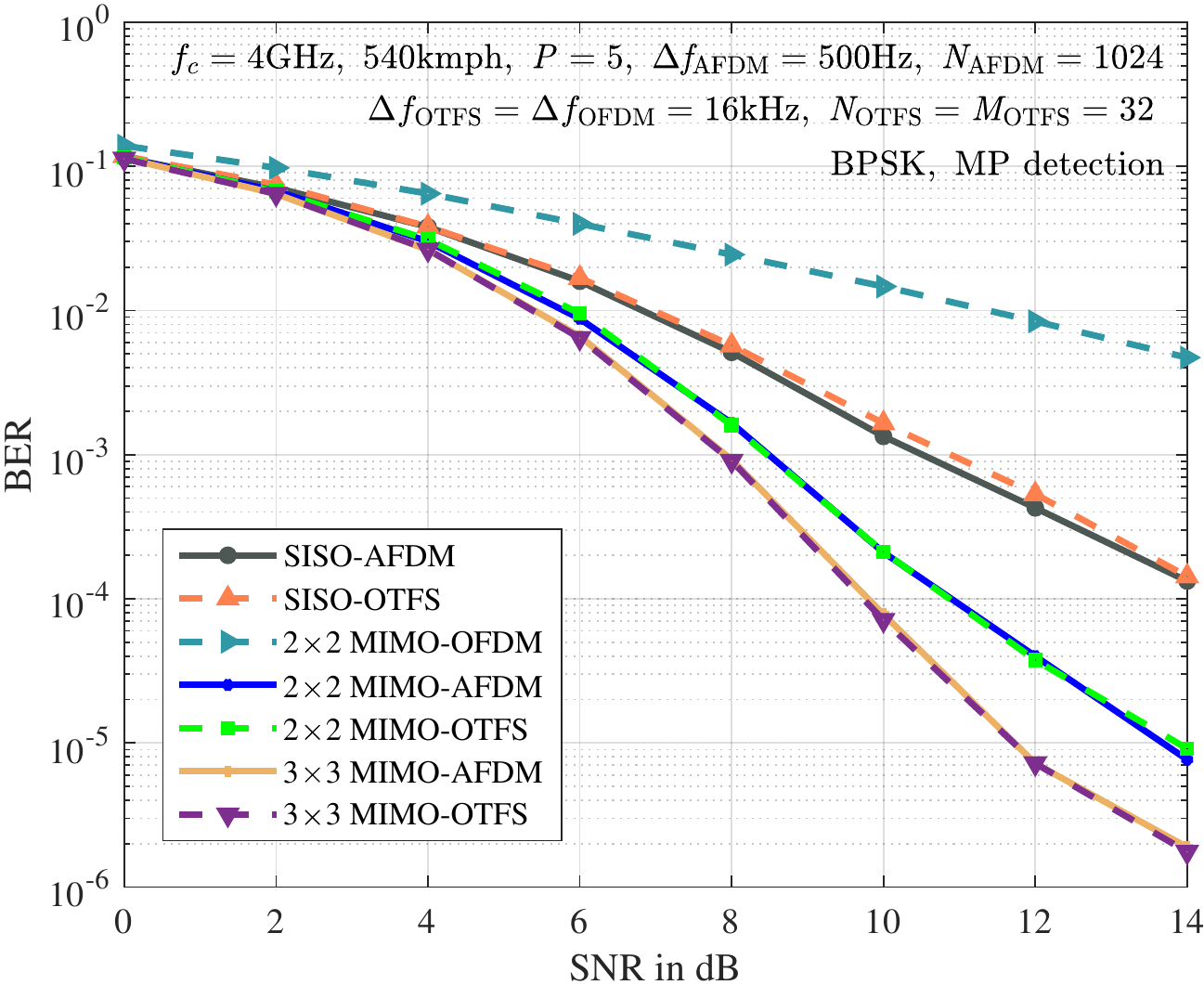}
		\vspace{-1.5em}
		\caption{BER comparisons among SISO-AFDM, SISO-OTFS, MIMO-OFDM, MIMO-AFDM and MIMO-OTFS systems with integer Doppler and $N=1024$. The normalized delay-Doppler profiles of five paths: (0, 0), (1, 1), (2, 2), (3, 3), (4, 4).}
		\label{sci3-5}
	\end{minipage}
	\quad
	\begin{minipage}[t]{0.45\textwidth}
		\centering
		\includegraphics[scale=0.56]{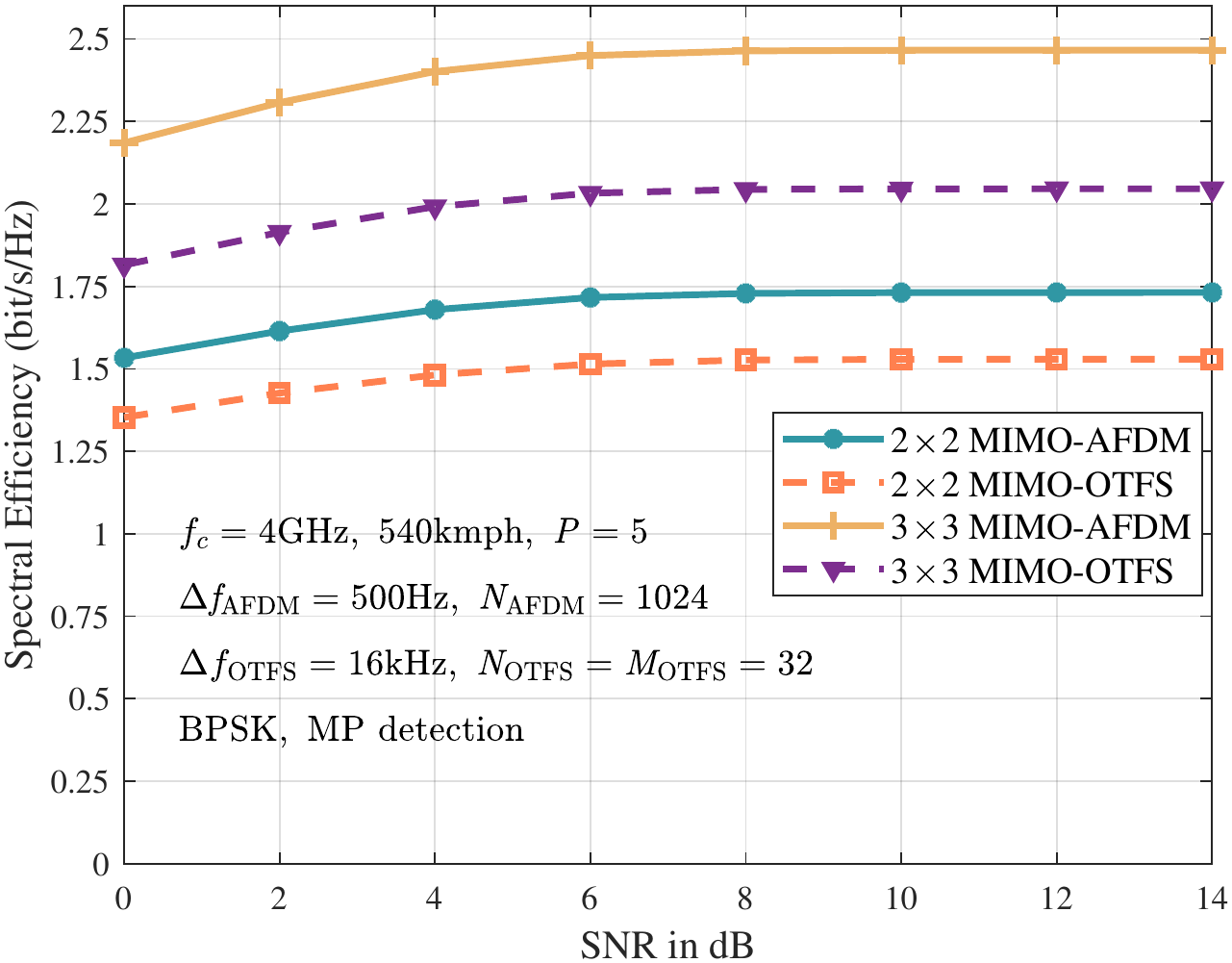}
		\vspace{-1.5em}
		\caption{Spectral efficiency comparison between the $2\times2$ MIMO-AFDM and the $2\times2$ MIMO-OTFS systems in Fig. \ref{sci3-5}.}
		\label{sci5-4-1}
	\end{minipage}
	\vspace{-1.3em}
\end{figure}

Fig. \ref{sci3-5} shows the BER performance comparisons among SISO-AFDM, SISO-OTFS, MIMO-OFDM, MIMO-AFDM, and MIMO-OTFS systems with integer Doppler.
Recalling that an OTFS frame with duration $N_{\text{OTFS}}T_{\text{OTFS}}$ (seconds) and bandwidth $M_{\text{OTFS}} \Delta f_{\text{OTFS}}$ (Hz) transmits $N_{\text{OTFS}}M_{\text{OTFS}}$ symbols, where $T_{\text{OTFS}}$ and $\Delta f_{\text{OTFS}}$ represent the time and frequency domain sample intervals, respectively, and satisfy $T_{\text{OTFS}}\Delta f_{\text{OTFS}}=1$; $N_{\text{OTFS}}$ and $M_{\text{OTFS}}$ denote the number of samples in the time-frequency plane (or Doppler-delay plane). We adopt  
$f_{c}=4$ GHz, $\Delta f_{\text{AFDM}}=500$ Hz, $N_{\text{AFDM}}=1024$, $\Delta f_{\text{OTFS}}=\Delta f_{\text{OFDM}}=16$ kHz, $N_{\text{OTFS}}= M_{\text{OTFS}}=32$ to ensure the same resources are occupied, i.e., bandwidth $B_{\text{AFDM}}=N_{\text{AFDM}}\Delta f_{\text{AFDM}}=M_{\text{OTFS}}\Delta f_{\text{OTFS}}=B_{\text{OTFS}}=512$ kHz. $P=5$ paths with BPSK and MP detection are applied. We can observe that the $2 \times 2$ MIMO-AFDM system outperforms the $2 \times 2$ MIMO-OFDM system significantly. This is because MIMO-OFDM suffers from severe inter-carrier interference caused by the heavy Doppler shifts. Moreover, owing to the equivalent delay-Doppler channel representation in the DAFT domain, AFDM establishes nearly the same BER performance as OTFS in SISO and MIMO configurations with large frame sizes\footnote{It is worth mentioning that, although the diversity order of uncoded OTFS is proven to be one, it enjoys nearly full diversity in the finite SNR region when the frame size $N_{\text{OTFS}}M_{\text{OTFS}}$ is sufficiently large \cite{bb4}. Therefore, the above observations that AFDM systems show almost the same BER performance as OTFS systems do not conflict with the conclusion that MIMO-AFDM can achieve full diversity.}. However, considering the lower overhead of MIMO-AFDM system when conducting the pilot-aided channel estimation, MIMO-AFDM will enjoy higher spectral efficiency than MIMO-OTFS. For example, according to (\ref{eq4-101}) and (\ref{eq7-28-1}), the pilot and guard overhead of $2\times2$ MIMO-AFDM for channel estimation is 134 slots, corresponding to 13.08\% of the entire AFDM frame, while the $2\times2$ MIMO-OTFS counterpart is 238 slots, corresponding to 23.24\% of the entire OTFS frame. This difference translates into a huge spectral efficiency gap between MIMO-AFDM and MIMO-OTFS, as shown in Fig. \ref{sci5-4-1}, where the values of spectral efficiency were calculated from the
BER values in Fig. \ref{sci3-5}.

\vspace{-0.5em}
\subsection{Imperfect CSI}
\label{sec4-3}
We next apply the EPA-DR scheme for channel estimation. $P=4$ paths with normalized delay profile of  $[0,\ 0,\ 1,\ 2]$ are adopted to cover all the cases in Table \ref{tabel7-31-1}. The Doppler shift of each path is generated by using Jakes' formula, i.e., $\nu_{i}=\nu_{\max } \cos \left(\theta_{i}\right)$, where $\theta_{i}$ is uniformly distributed over $[-\pi, \pi]$, $\nu_{\max}=2$ corresponds to a maximum Doppler shift of 2000 Hz and a maximum UE speed of 540 kmph. The pilot signal for SNR is denoted as SNRp, while the received data signal for SNR is denoted as SNRd. Other
parameters considered for the simulation are provided in Table \ref{t3}. 

\begin{table}[ht]
	\vspace{-2em}
	\renewcommand\arraystretch{0.65}
	\centering
	\caption{Simulation Parameters for Imperfect CSI}
	\label{t3}
	\vspace{-1.2em}
	\begin{tabular}{|c|c|}
		\hline
		\textbf{Parameter}& \textbf{Value} \\ 
		\hline
		Carrier frequency $f_{c}$ (GHz)& 4 \\ 
		\hline
		Subcarrier spacing $\Delta f_{\text{AFDM}}$ (kHz)& 1 \\
		\hline
		Number of subcarrier (frame size) $N_{\text{AFDM}}$ & 1024 \\			
		\hline
		Bandwidth $B_{\text{AFDM}}$ (kHz) & 1024 \\	
		\hline
		Maximum UE speed (kmph)&  540
		 \\ 
		\hline
		 Maximum Doppler shift (kHz)&  2 \\  		
		\hline
		Number of paths ($P$) & 4 \\ 
		\hline	
		Modulation scheme&
		\tabincell{c}{BPSK, 4-QAM,\\ and 16-QAM}
		  \\ 
		\hline
		MIMO configuration &  $2\times2$  \\ 
		\hline
		Detector& MP \\ 
		\hline				
	\end{tabular}
\vspace{-1.3em}
\end{table}

\begin{figure}[h]
	\vspace{-1.3em}
	\begin{minipage}[t]{0.5\textwidth}
		\centering
		\includegraphics[scale=0.5]{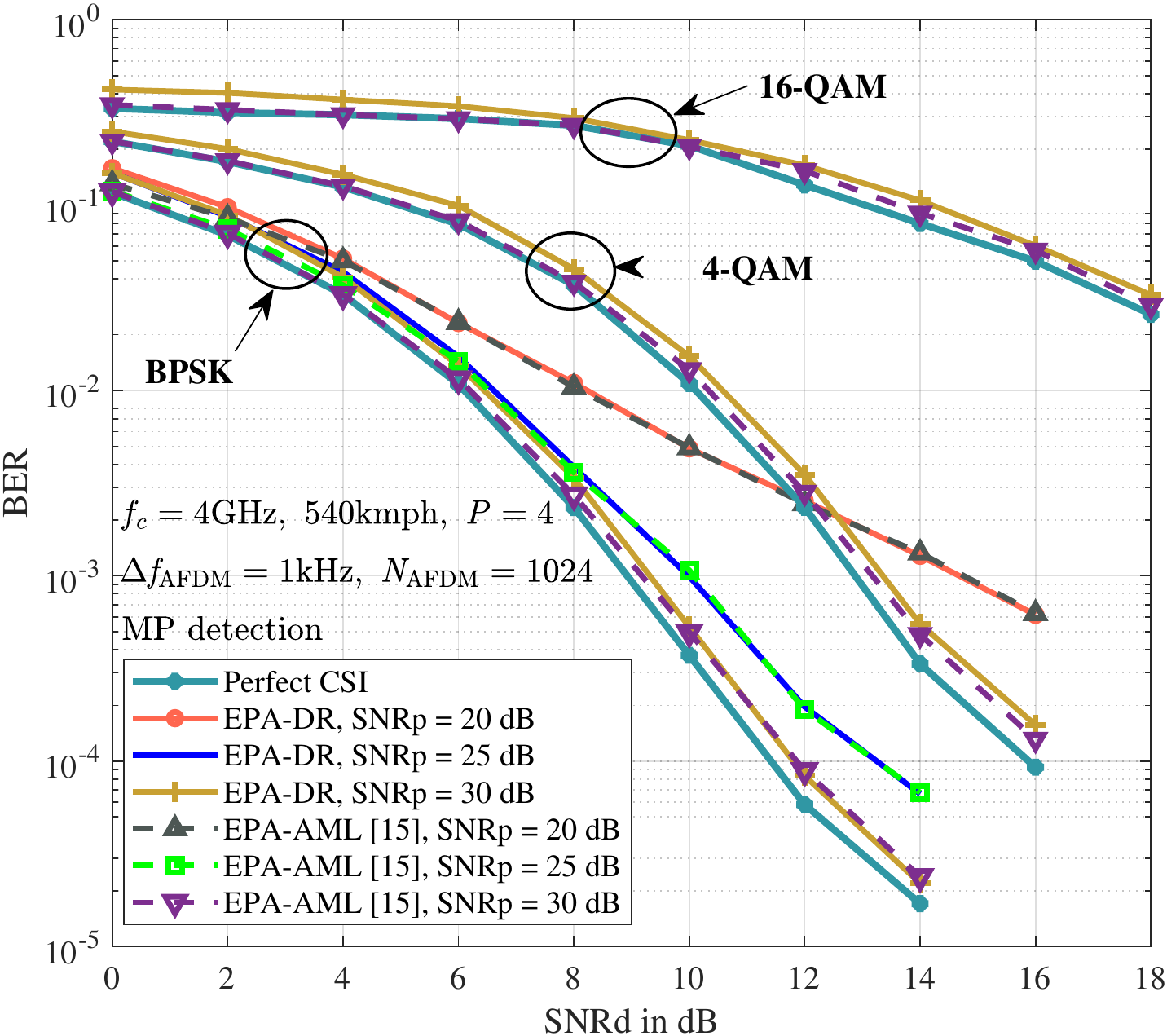}
		\vspace{-1.8em}
		\caption{BER versus SNRd of $2\times2$ MIMO-AFDM system \\  with  different SNRp, integer Doppler.}
		\label{5-5}
	\end{minipage}
	\quad
	\begin{minipage}[t]{0.5\textwidth}
		\centering
		\includegraphics[scale=0.57]{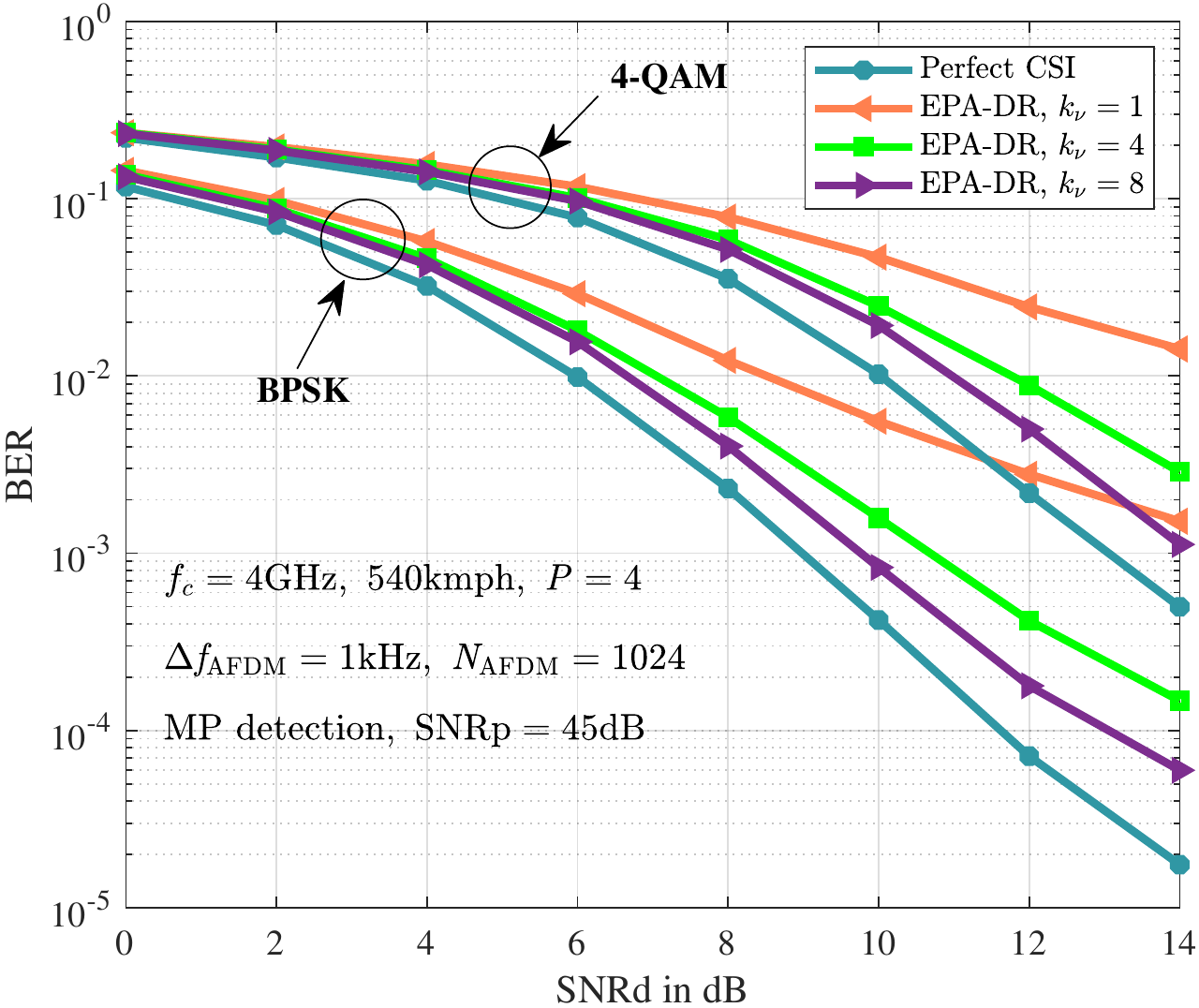}
		\vspace{-1.8em}
		\caption{BER versus SNRd of $2\times2$ MIMO-AFDM system with different spacing factors $k_{\nu}$, SNRp $=45$ dB, fractional Doppler.}
		\label{5-7}
	\end{minipage}
	\vspace{-2em}
\end{figure}

Fig. \ref{5-5} shows the BER versus SNRd of MIMO-AFDM with different SNRp. Integer Doppler is considered. In this case, there is no IDoI, IDI, and IPI among the received pilot symbols and thus the EPA-AML scheme \cite{bb6} is feasible. We can observe that the BER performance enhances as SNRp increases. This is because the only error is the AWGN, which is normalized by $x_{t}^{\text{pilot}}$ in \textbf{Step 2} of EPA-DR. Therefore, the larger energy of the pilot, the more accurate the estimated $\mathbf{\hat{H}}_{r,t}$. Moreover, when SNRp reaches 30 dB, the BER of MIMO-AFDM system using the estimated CSI shows only marginal degradation compared to that with Perfect CSI, which validates the effectiveness of the EPA-DR scheme. In addition, we can notice that, with much lower computation complexity, EPA-DR delivers nearly the same channel estimation performance as EPA-AML.

We next investigate the effect of spacing factor $k_{\nu}$ on the BER performance of MIMO-AFDM system with fractional Doppler. In this case, the EPA-AML suffers from serious IDoI and hence is impracticable. Therefore, we simulate three $k_{\nu}$ values ($k_{\nu}$ = 1, 4, and 8) with the proposed EPA-DR scheme. We can observe from Fig. \ref{5-7} that, as $k_{\nu}$ increases, the BER performance improves. This is because the protection bands between two contiguous delay blocks in the effective channel matrix are widened, as shown in Fig. \ref{sci4-3}, making the IDI, IPI, and IPDI smaller. However, according to (\ref{eq4-101}), as $k_{\nu}$ increases, the pilot and guard overhead $O_{\text{MIMO-AFDM}}$ will increase correspondingly (the $O_{\text{MIMO-AFDM}}$ of the three used $k_{\nu}$ values are 62, 116, and 188, respectively, corresponding to 6.05\%, 11.33\%, and 18.40\% of the entire AFDM frame), leading to fewer data symbols can be transmitted in each transmit antenna. We can also notice that compared to the integer Doppler case in Fig. \ref{5-5}, a higher SNRp is required for the fractional Doppler case.

\begin{figure}[h]
	\vspace{-1.3em}
	\begin{minipage}[t]{0.5\textwidth}
		\centering
		\includegraphics[scale=0.51]{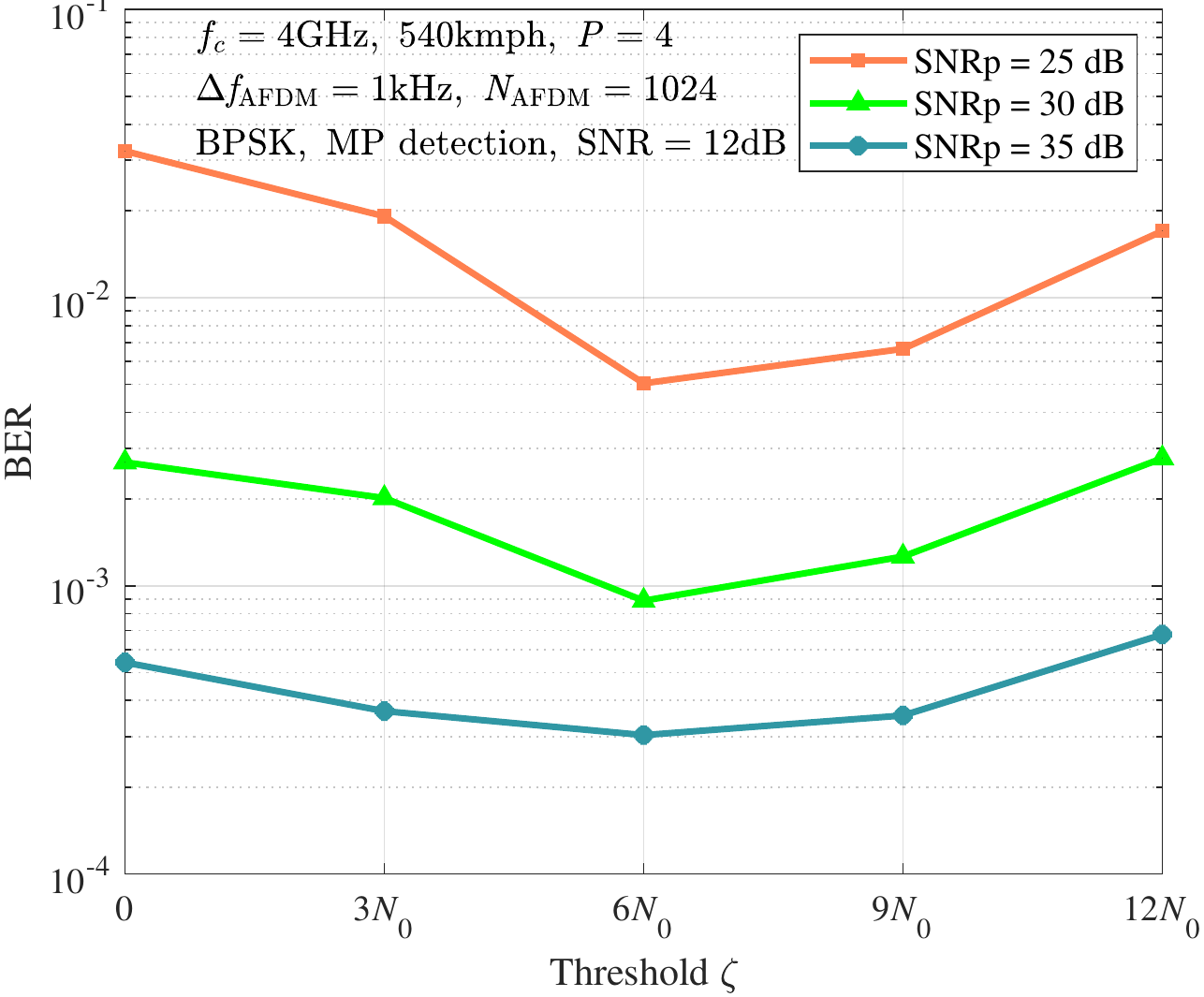}
		\vspace{-1.8em}
		\caption{BER versus different thresholds of $2\times2$ MIMO-AFDM system, fractional Doppler, $k_{\nu}=8$.}
		\label{5-6}
	\end{minipage}
	\quad
	\begin{minipage}[t]{0.45\textwidth}
		\centering
		\includegraphics[scale=0.52]{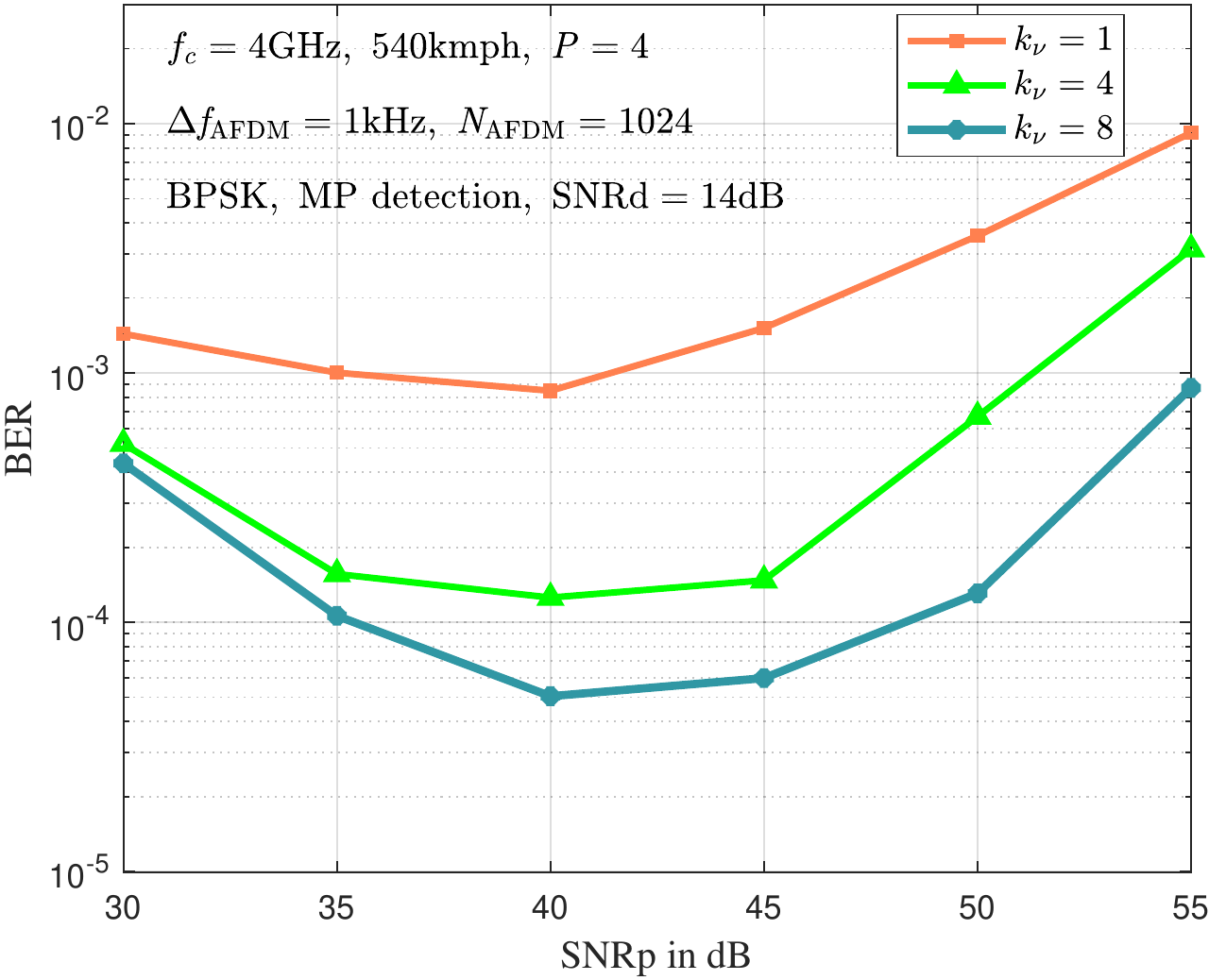}
		\vspace{-1.8em}
		\caption{BER versus SNRp of $2\times2$ MIMO-AFDM system with different spacing factors $k_{\nu}$, SNRd $=14$ dB, fractional Doppler.}
		\label{5-8}
		\vspace{-0.5em}
	\end{minipage}
	\vspace{-1.8em}
\end{figure}

Fig. \ref{5-6} shows the BER versus different thresholds of MIMO-AFDM system with fractional Doppler. We set the thresholds by taking  $N_{0}$ (the variance of noise) as a reference. Three different SNRp values are adopted. Setting $\zeta=0$ represents that we do not conduct the threshold-based magnitude detection in \textbf{Step 2} of EPA-DR. We can observe that there exists an optimal threshold, which is around $6N_{0}$. According to (\ref{eq4-37}),
if $\zeta$ is set higher than the optimal threshold, the CSI in the received pilot symbols with small energy will be omitted inappropriately; if $\zeta$ is set lower than the optimal threshold, the interference of noise with large energy will be reserved. Both of the cases will deteriorate the accuracy of EPA-DR channel estimation, followed by the detection performance degradation. Moreover, we find that with SNRp increasing, the BER performance is less sensitive to the threshold variation. This is because with the increase of pilot symbols' energy, the IDI, IPI will dominate the overall interference in EPA-DR gradually, as analyzed in (\ref{eq4-56}) and Remark 4.

Fig. \ref{5-8} shows the BER versus SNRp of MIMO-AFDM in the case of fractional Doppler. We can observe that, as the SNRp increases, the BER performance first enhances and then degrades. This is because when SNRp is smaller than 40 dB, the noise plays the major role of interference in channel estimation and can be suppressed by enlarging the energy of the pilot. However, according to (\ref{eq4-57-10}), the IPDI in the received
data symbols will be magnified at the same time, making signal detection more difficult even if a more accurate channel estimate is obtained. Therefore, there is a tradeoff between the pilot-aided channel estimation accuracy and the
pilot-data symbols separability at the receiver in AFDM with fractional Doppler, which can be interpreted as the cost of compressing the two-dimensional delay-Doppler domain into
the one-dimensional DAFT domain.

\vspace{-0.5em}
\section{Conclusion}
\label{sec6}
In this paper, we investigate the
channel estimation in MIMO-AFDM  with specific attention to fractional Doppler shifts. With ideal CSI, MIMO-AFDM is shown to achieve full diversity and establishes a BER performance similar to MIMO-OTFS and superior to MIMO-OFDM significantly, while showing a great advantage in spectral efficiency over MIMO-OTFS. Next, by exploring the unique diagonal reconstructability of the AFDM subchannel matrix, we propose a low-complexity channel estimation scheme for MIMO-AFDM to eliminate the serious IDoI induced by fractional Doppler shifts. Finally, we propose an orthogonal resource allocation scheme for AFDMA system. Simulation results show that the BER performance of MIMO-AFDM applying EPA-DR channel estimation is very close to the case with ideal CSI. In the future, space-time coding can be investigated to acquire transmit diversity in MIMO-AFDM system with the proposed EPA-DR channel estimation method \cite{bb23.3.6.1}. 

\vspace{-1.2em}
{\appendix[Proof of  theorem 1]
	\vspace{-0.5em}
	\label{APP1}
	For ease of derivations, we assume the Doppler shifts as integers. Simulation results given in Section \ref{secResults_idealCSI} show that the derived diversity order also holds for the fractional Doppler case.
	
	Considering there are $P$ propagation paths between all  pairs of RAs and TAs, and the fractional parts of all the associated Doppler shifts are zero ($\beta_{i}= 0$ and $k_{\nu}=0$), then the spreading factor $\mathcal{F}(l_{i},\nu_{i}, m, m')$ in (\ref{eq2-11-3}) satisfies
	\begin{equation}
		\vspace{-0.2em}
		\begin{aligned}
			\mathcal{F}(l_{i},\nu_{i}, m, m') 
			&= \frac{e^{j 2 \pi\left(m+\operatorname{ind}_{i}-m'\right)}-1}{e^{j \frac{2 \pi}{N}\left(m+\operatorname{ind}_{i}-m'\right)}-1} = \left\{\begin{array}{ll}
				N, & m'=\left(m+\operatorname{ind}_{i}\right)_{N} \\
				0, &  otherwise. 
			\end{array}\right.
		\end{aligned}
		\label{eq3-18-2}
	\end{equation}
	Hence, the input-output relationship of SISO-AFDM in  (\ref{eq2-9}) can be simplified as
	\vspace{-0.3em}
	\begin{equation}
		\label{eq2-17-2}
		y[m] =\sum_{i=1}^{P} h_{i} \mathcal{C}(l_{i}, m, m') x[m']+w[m]
	\end{equation}
	where $ m \in [0,N-1]$, $m'=\left(m+\operatorname{ind}_{i}\right)_{N}$. Therefore, there are $P$ non-zero elements in each row and column of $\mathbf{H}_{r,t}$, $\forall \ r \in [1,N_{r}], t \in [1,N_{t}]$. Equation (\ref{eq2-12-2}) can then be presented in an alternate way as \cite{bb6}
	\begin{equation}
		\mathbf{y} = \sum_{i=1}^{P} h_{i} \mathbf{H}_{i} \mathbf{x}+\mathbf{w}=\boldsymbol{\Phi}(\mathbf{x}) \mathbf{h}+\mathbf{w}
	\end{equation}
	where $\mathbf{\Phi}(\mathbf{x}) = \left[\mathbf{H}_{1} \mathbf{x}, \ \mathbf{H}_{2} \mathbf{x}, \ \ldots, \  \mathbf{H}_{P} \mathbf{x} \ \right]\in \mathbb{C}^{N \times P}$, channel gain vector $\mathbf{h} = [h_{1}, h_{2}, \ldots, h_{P}]^{T} \in \mathbb{C}^{P\times 1}$.

	Similarly, the input-output relationship of MIMO-AFDM in  (\ref{eq2-111}) can be simplified as
	\vspace{-0.2em}
	\begin{equation}
		\label{eq2-15}
		y_{r}[m] =\sum_{t=1}^{N_{t}}\sum_{i=1}^{P} h_{i}^{[r,t]}\mathcal{C}(l_{i}, m, m')x_{t}[m']+w_{r}[m]
	\end{equation}
	where  $m'=\left(m+\operatorname{ind}_{i}\right)_{N}$.  From (\ref{eq2-15}) we can see that each row of $\mathbf{H}_{\text{MIMO}}$ has only $PN_{t}$ non-zero elements. Therefore, the vectorized input-output relationship in (\ref{eq.8}) can also be presented as 
	\vspace{-0.5em}
	\begin{equation}
		\mathbf{Y} = \mathbf{\tilde{\Phi}}(\mathbf{X})\mathbf{\tilde{h}}  + \mathbf{W}
		\vspace{-0.5em}
	\end{equation}
	where received symbol matrix $\mathbf{Y} = [\mathbf{y}_{1}, \mathbf{y}_{2}, \ldots, \mathbf{y}_{N_{r}}]$ is an $N \times N_{r}$ matrix  whose $r$-th column is the received symbol vector of the $r$-th RA, transmitted symbol matrix $\mathbf{X} = [\mathbf{x}_{1}, \mathbf{x}_{2}, \ldots, \mathbf{x}_{N_{t}}]$ is an $N \times N_{t}$ matrix  whose $t$-th column is the transmitted symbol vector of the $t$-th TA, $\mathbf{\tilde{\Phi}}(\mathbf{X})$ is an $N \times PN_{t}$ concatenated matrix with a definition of 
	\vspace{-0.5em}
	\begin{equation}
		\mathbf{\tilde{\Phi}}(\mathbf{X}) = \left[ \ \mathbf{\Phi}(\mathbf{x}_{1}), \ \mathbf{\Phi}(\mathbf{x}_{2}), \ \ldots, \ \mathbf{\Phi}(\mathbf{x}_{N_{t}}) \ \right]
		\vspace{-0.3em}
	\end{equation}
	$\mathbf{\tilde{h}} \in \mathbb{C}^{PN_{t} \times N_{r}}$ is channel gain matrix with a definition of 
	\vspace{-0.1em}
	\begin{equation}
		\mathbf{\tilde{h}}=\left[\begin{array}{cccc}
			\mathbf{h}_{1,1} &  \ldots & \mathbf{h}_{N_{r} ,1} \\
			\vdots &  \ddots & \vdots \\
			\mathbf{h}_{1, N_{t}} &  \ldots & \mathbf{h}_{N_{r}, N_{t}}
		\end{array}\right]
	\vspace{-0.1em}
	\end{equation}
	where $\mathbf{h}_{r, t} \in \mathbb{C}^{P\times 1}$ denotes the channel gain vector between the $r$-th RA and $t$-th TA, and its elements $h_{i}^{r,t}$ are assumed to follow the distribution of $\mathcal{C N}(0,1 / P)$ (uniform scattering profile), $\mathbf{W} \in \mathbb{C}^{N\times N_{r}}$ is the noise matrix.
	
	For convenience, we normalize the transmit symbol matrix $\mathbf{X}$ so that the signal-to-noise ratio (SNR) at each receive antenna is $\frac{1}{N_{o}}$. Let $\mathbf{X}_{i}$ and $\mathbf{X}_{j}$ be two transmit symbol matrices. Assuming perfect CSI and maximum likelihood (ML) detection at the receiver, the probability of transmitting the symbol matrix $\mathbf{X}_{i}$ and deciding in favor of $\mathbf{X}_{j}$ at the receiver is the pairwise error probability (PEP) between $\mathbf{X}_{i}$ and $\mathbf{X}_{j}$, which can be expressed as \cite{bb14}
	\begin{equation}
		P\left(\mathbf{X}_{i} \rightarrow \mathbf{X}_{j} \mid \mathbf{\tilde{h}}, \mathbf{X}_{i}\right)=Q\left(\sqrt{\frac{\left\|\left(
				\mathbf{\tilde{\Phi}}(\mathbf{X}_{i})-\mathbf{\tilde{\Phi}}(\mathbf{X}_{j})\right)\mathbf{\tilde{h}}\right\|^{2}}{2 N_{0}}}\right).
	\end{equation}
	The PEP averaged over the channel statistics can be
	denoted as
	\begin{equation}
		P\left(\mathbf{X}_{i} \rightarrow \mathbf{X}_{j} \right)=\mathbb{E}\left[Q\left(\sqrt{\frac{\left\|\left(\mathbf{\tilde{\Phi}}(\mathbf{X}_{i})-\mathbf{\tilde{\Phi}}(\mathbf{X}_{j})\right)\mathbf{\tilde{h}}\right\|^{2}}{2 N_{0}}}\right)\right].
		\label{eq3-22}
	\end{equation}
	Define difference matrix as $\boldsymbol{\delta}^{(i, j)}\triangleq \mathbf{X}_{i}-\mathbf{X}_{j}$, where the $t$-th column of $\boldsymbol{\delta}^{(i, j)}$ is  $\boldsymbol{\delta}_{t}^{(i, j)} = \mathbf{x}_{t}^{(i)}-\mathbf{x}_{t}^{(j)}$, which denotes the difference of two transmitted symbols at the $t$-th TA. Since ${\tilde{\mathbf{\Phi}}}$ is a linear operator, we have 
	\begin{equation}
		P\left(\mathbf{X}_{i} \rightarrow \mathbf{X}_{j} \right)=\mathbb{E}\left[Q\left(\sqrt{\frac{\left\|\mathbf{\tilde{\Phi}}(\boldsymbol{\delta}^{(i, j)})\mathbf{\tilde{h}}\right\|^{2}}{2 N_{0}}}\right)\right].
		\label{eq3-22}
	\end{equation}
	Using Chernoff bound and the fact that each TA transmits independent AFDM symbols, an upper bound on the PEP in (\ref{eq3-22}) can be obtained as \cite{bb14}
	\begin{equation}
		P\left(\mathbf{X}_{i} \rightarrow \mathbf{X}_{j}\right) \leq\left(\prod_{l=1}^{k} \frac{1}{1+\frac{ \lambda_{t, l}^{2}}{4 P N_{0}}}\right)^{N_{r}}
		\label{eq3-23}
	\end{equation}
	where $\lambda_{t, l}$ is the $l$-th singular value of the matrix $\mathbf{\Phi}(\boldsymbol{\delta}_{t}^{(i, j)})$ ($t \in 1,2, \ldots, N_{t}$ ) and $k$ is the rank of $\mathbf{\Phi}(\boldsymbol{\delta}_{t}^{(i, j)})$. At high SNR regime, (\ref{eq3-23}) can be further simplified as 
	\begin{equation}
		P\left(\mathbf{X}_{i} \rightarrow \mathbf{X}_{j}\right) \leq\frac{1}{N_{o}^{kN_{r}}}\left(\prod_{l=1}^{k} \frac{\lambda_{t, l}^{2}}{4 P}\right)^{-N_{r}}.
		\label{eq3-25}
	\end{equation}
	We can observe from (\ref{eq3-25}) that the exponent of the SNR term $\frac{1}{N_{o}}$ is $kN_{r}$, and the overall bit error ratio (BER) is dominated by the PEP with the minimum value of $k$, for all $i, j, i \neq j$. Therefore, the diversity order of MIMO-AFDM, denoted by $\rho$, is given by
	\vspace{-0.3em}
	\begin{equation}
		\rho = N_{r} \cdot \min _{i, j, i \neq j} \operatorname{rank}\left(\mathbf{\Phi}(\boldsymbol{\delta}_{t}^{(i, j)})\right).
		\vspace{-0.3em}
	\end{equation}
	In \cite{bb6} (Appendix A), the term $\min _{i, j, i \neq j} \operatorname{rank}\left(\mathbf{\Phi}(\boldsymbol{\delta}_{t}^{(i, j)})\right)$ is proven to be $P$ whenever the AFDM parameters $c_{1}$ and $c_{2}$ are tuned as described in Remark 1, and the number of subcarriers $N$ satisfies
	\begin{equation}
		N\geq \left(l_{\max }+1\right)\left(2 \alpha_{\max }+1\right). 
		\label{eq3-27}
	\end{equation}
	In practice, $N \gg \left(l_{\max }+1\right)\left(2 \alpha_{\max }+1\right)$, thus the diversity order of MIMO-AFDM is $PN_{r}$, i.e., MIMO-AFDM achieves full diversity in doubly selective channels. This completes the proof of Theorem 1.}

\vfill
\end{document}